\documentclass[aps,prb,twocolumn,groupedaddress,showpacs,floatfix,superscriptaddress]{revtex4}
\usepackage{epsfig}
\usepackage{amsmath,amssymb}
\usepackage{graphicx}
\usepackage[dvipsnames,usenames]{color}
\usepackage[normalem]{ulem}
\emergencystretch=\maxdimen
\begin{document}

\title{Geometry Dependence of the Sign Problem}

\author{V. I.~Iglovikov}
\affiliation{Department of Physics, University of California, Davis, California 95616, USA}
\author{E. Khatami}
\affiliation{Department of Physics and Astronomy, San Jose State University, San Jose, CA 95192, USA}
\author{R. T. Scalettar}
\affiliation{Department of Physics, University of California, Davis, California 95616, USA}

\begin{abstract}
The sign problem is the fundamental limitation to quantum Monte Carlo
simulations of the statistical mechanics of interacting fermions.
Determinant quantum Monte Carlo (DQMC) is
one of the leading methods to study lattice models such as the Hubbard
Hamiltonian, which describe strongly correlated phenomena including
magnetism, metal-insulator transitions, and (possibly) exotic
superconductivity. Here, we provide a comprehensive dataset on the
geometry dependence of the DQMC sign problem for different densities,
interaction strengths, temperatures, and spatial lattice sizes. We
supplement these data with several observations concerning general trends
in the data, including the dependence on spatial volume and how this can
be probed by examining decoupled clusters, the scaling of the sign in
the vicinity of a particle-hole symmetric point, and the correlation
between the total sign and the signs for the individual spin species.
\end{abstract}

\maketitle

\section{Introduction}

Monte Carlo simulations of classical systems have generated very precise
information about phases and phase transitions in statistical mechanics.
One dramatic example of the power of the methodology is that of the
Ising model, where the transition temperature on a cubic lattice is now
known to six decimal places~\cite{gupta96} and critical exponents have
been evaluated to four decimal places.~\cite{hasenbusch99}  A roughly
similar situation holds for unfrustrated quantum spin and boson systems.
For example, in the Bose-Hubbard Hamiltonian,~\cite{fisher89} the
critical interaction strengths at a density of one boson per site for
the superfluid to Mott insulator transition in the ground state, are
available~\cite{capogrossosansome07} to an accuracy of better than one
part in $10^3$.  Spatial lattice sizes are somewhat smaller than for
classical problems since writing the partition function as a path
integral introduces an additional `imaginary time' dimension, but are
nevertheless quite large, e.g.,~up to $10^4$ sites for the Bose-Hubbard
example cited above.

Fermions (in more than one dimension) are more challenging for two reasons.  First, the fermionic action is non-local: the Boltzmann weight typically takes the form of a determinant.  Thus, updating all the degrees of freedom has a computation time which scales as a nonlinear power of the system size $N$. The cost of a method like determinant quantum Monte Carlo (DQMC)~\cite{blankenbecler81} scales as $N^3$, as opposed to the naive linear in $N$ scaling (ignoring such complications as critical slowing down) in many classical and quantum spin/boson applications.  Second, and far worse, there is no guarantee the sign of the determinant, which is used as the probability, is positive.  Although one can formally use the absolute value as a weight, and include the sign in the measurement, in practice one ends up evaluating the ratio of two numbers, which becomes dominated by statistical error at low temperatures as they both become very small. This situation is known as the ``fermion sign problem",~\cite{loh90, dos_santos2003} whose solution is conjectured to be NP-hard.~\cite{troyer05}  At present, therefore, there is no known method of accessing the low temperature properties of Hamiltonians like the Fermi-Hubbard model with QMC.

Fortunately, there are some situations where the sign problem is not manifest.  For example, the Boltzmann weight often takes the form of the product of  two determinants, one for each of the two electron spin species, and it can happen that the signs of the individual determinants perfectly match, so that the Boltzmann weight remains positive.  This occurs in the complete parameter range of the Hubbard model with attractive interactions, enabling a study of superconductivity and charge density wave physics.
It also occurs in the Hubbard model with repulsive interactions on bipartite lattices in the limit of average one particle per lattice site (half-filling), so that the Mott transition and long-range antiferromagnetism can be explored.  Other instances of situations where the sign problem is absent are mentioned in Sec.~IIC.
However, many of  the most interesting questions concerning strong correlation physics remain inaccessible, most notably the question of whether the two-dimensional (2D) square lattice repulsive Hubbard Hamiltonian has a low temperature $d$-wave superconducting transition, so that it would provide a good description of cuprate superconductivity.~\cite{scalapino94}

We have two main goals in this manuscript.  The first is to present a set of data for the sign problem in DQMC for different geometries. These include bipartite [one-dimensional chain, ladder, 2D square, and three-dimensional (3D) cubic, Lieb, honeycomb and the 1/5-depleted square] lattices, and non-bipartite Kagome and triangular lattices.  We consider DQMC because it is a powerful and widely utilized approach to the correlated electron problem.  Our second goal is to discern trends in the DQMC sign problem.  We will consider, for example, several new issues: how the sign depends on the spatial lattice size, the scaling of the sign in the vicinity of particle-hole symmetric (PHS) points, and the `entanglement' of the sign as probed by the consideration of coupled cluster geometries. Although statements concerning the first of these points, the spatial size dependence, have been made in the literature, numerical data are rather scanty, owing to the computational limitations existing in the initial investigations. Specifically, the data in Refs.~\onlinecite{dos_santos2003} and \onlinecite{white89} were restricted to 4x4, 6x6, and 8x8 lattices.  As a consequence, the scaling regime was not reached for many of the parameter sets.  For example, the average sign sometimes {\it increases} as the system size grows, rather than decreasing. This situation is rectified here. The lattices we study also have various unique features, such as nesting of the Fermi surface, van-Hove singularities in the density of states $N(\omega) = \sum_k \, \delta(\omega - \epsilon_k)$, and flat bands, whose possible effects on the sign problem we will examine.

It is worth noting what we will {\it not} cover: There are a number of methods which are closely related to DQMC in that they involve a Hubbard-Stratonovich decoupling of the interaction, and a Boltzmann weight built from fermion determinants.  These include impurity algorithms,~\cite{hirsch86} as well as the dynamical mean field theory (DMFT)~\cite{vollhardt93,georges96,dmft_3} and its cluster extensions, the dynamical cluster approximation (DCA),~\cite{maier05} and the cellular DMFT.~\cite{kyung06} A strength of some of these methods is that the sign problem is greatly mitigated relative to DQMC, at least if the cluster size is not too large.  This is true even if the bath degrees of freedom are discretized.~\cite{khatami09}  Also closely related to DQMC are zero-temperature algorithms which use $e^{-\beta \hat H}$ as a projection operator on a trial wave function.~\cite{sorella89,white89} Constraints can be introduced in these ground state methods to eliminate the sign problem, at the expense of systematic errors in the solution.~\cite{zhang97} Despite their relations to DQMC, we will not discuss these approaches here. Similarly, within DQMC itself there are different choices of the manner in which the Hubbard-Stratonovich field is decoupled.~\cite{buendia86,tremblay92,batrouni90} Here, we base our calculations on only the ``density decoupling", described in Sec. IIB. Though we do not explicitly consider the above related approaches, we expect that some of our results and general analysis may have applications to them as well.

The remainder of this paper is organized as follows: In Sec.~II we review the Hubbard Hamiltonian and the basic formulation of DQMC, followed by some general, well-known properties of the resulting sign problem.  In Sec.~III we record values for the average sign for different lattice geometries.  Section IV examines general patterns in this data.  Finally, Sec.~V contains concluding remarks.

\section{General Considerations Concerning the Sign Problem}

\subsection{The Hubbard Hamiltonian}

Our focus is on the single-band Hubbard Hamiltonian,
\begin{align}
\hat H = &- \sum_{\langle ij \rangle \sigma}
t_{ij}\big( \,
c_{i \sigma}^{\dagger}   \, c_{j \sigma}^{\phantom{\dagger}}
+
c_{j \sigma}^{\dagger}   \, c_{i \sigma}^{\phantom{\dagger}}
 \, \big)
 - \mu \sum_{i\sigma} n_{i \sigma}
\nonumber \\
 &+ U \sum_{i}
 \left(\, n_{i\uparrow} -\frac12 \, \right)
 \left(\, n_{i \downarrow } -\frac12 \, \right)
\label{eq:ham}
\end{align}
Here $c_{i\sigma}^{\dagger}(c_{i\sigma}^{\phantom{\dagger}})$ is the
creation (destruction) operator for a fermion with spin $\sigma$ on
site $i=1,2,\cdots N$ and $n_{i\sigma}^{\phantom{\dagger}}
=c_{i\sigma}^{\dagger}c_{i\sigma}^{\phantom{\dagger}}$ is the number
operator. $t_{ij}$ is the hopping
amplitude between nearest-neighbor sites $i$ and $j$, $U$ is the interaction strength, and
$\mu$ is the chemical potential. For geometries where there is
only one type of hopping, we set $t_{ij}=t=1$ as the unit of energy.
We will denote the first line of the Hamiltonian $\hat H$
in Eq.~(\ref{eq:ham}) by $\hat K$, and the second line by $\hat V$.
This latter term is written in PHS form.  (See Sec.~IIC.) The
different models considered in this paper are distinguished solely by
the geometry encoded in the near-neighbor designation $\langle ij
\rangle$ in the kinetic energy term.  Even with a common choice $t=1$,
different geometries have distinct bandwidths $W$, the spread of
eigenvalues of the $U=0$ (single particle) Hamiltonian.
Although it is sometimes the case~\cite{khatami14}
that using $W$ as the scale of kinetic
energy, rather than $t$, produces better comparisons across different
models, we did not find that to be useful here.
We retain the standard convention of normalizing to $t$.

There is, of course, much interest in generalizations of the Hubbard
Hamiltonian, e.g.,~to multiple bands, longer-range density-density
interactions, and Hund's rule type interactions.  Multiple bands, can in
fact be written in the form of Eq.~(\ref{eq:ham}), with the understanding
that the label $i$ incorporates both spatial and band indices.  Thus,
from the viewpoint of a DQMC simulation, setting up the 2D `periodic
Anderson model' (PAM) which has a square lattice of spatial sites and
two orbitals per site, is formally identical to a two layer geometry, in
which there is a single orbital on each site.
Thus, with the freedom to choose $U_i \, (t_{ij})$ to be
site/orbital (bond) dependent, Eq.~(\ref{eq:ham}) incorporates Hamiltonians
like the PAM.  Concerning intersite (interorbital) and Hund's rule
interactions, the sign problem is typically much worse than for an
on-site $U$ between fermions of different spin species only.  For
example, for a model of the CuO$_2$ planes of cuprate
superconductors,~\cite{scalettar91} it was found that the sign problem
restricted simulations to interactions $U_{pd} \lesssim 1$ at
temperatures where local spin correlations were seen to begin to
develop.  Like DQMC, Hund's rule interactions also present grave sign
problem difficulties in DMFT.~\cite{hunddmft}  We do not explicitly consider them
here.

\subsection{Determinant Quantum Monte Carlo}

The fundamental idea of DQMC~\cite{footnote1} is to take advantage of the fact that it is possible to compute analytically the trace of a product of the exponentials of quadratic forms of fermion creation and destruction operators.  If we denote the vector of creation operators
$(\, c_{1} ^{\dagger} \,,
 c_{2} ^{\dagger} \,,
 c_{3} ^{\dagger} \, \cdots \,
 c_{N} ^{\dagger} \,) $
by $\vec c^{\, \dagger}$, and $A_j$ are (symmetric) $N \times N$ matrices
of real numbers, then
\begin{align}
{\rm Tr} & \, (\, e^{\vec c^{\, \dagger} A_1 \vec c^{\phantom{\dagger}} }
e^{\vec c^{\, \dagger} A_2 \vec c^{\phantom{\dagger}} }
e^{\vec c^{\, \dagger} A_3 \vec c^{\phantom{\dagger}} }
\cdots
e^{\vec c^{\, \dagger} A_L \vec c^{\phantom{\dagger}} } \, )
\nonumber \\
&= {\rm det} \, ( \, I + B_1 B_2 B_3 \cdots B_L \, )
\label{eq:quadformidentity}
\end{align}
Here $B_i = e^{A_i}$.  It is important to emphasize that the
trace in the left hand side of Eq.~(\ref{eq:quadformidentity})
is over a $2^N$ dimensional Hilbert space of the fermionic operators
while the determinant on the right hand side is taken over
a real matrix of dimension $N$.

The interaction term in the Hubbard Hamiltonian Eq.~(\ref{eq:ham})
is not quadratic in the fermionic operators, but can be made so by
first discretizing the inverse temperature $\beta = L \tau$
and then employing the Trotter approximation~\cite{trotter59,suzuki76,fye86}
\begin{align}
{\rm Tr} \, e^{-\beta \hat H} & =
{\rm Tr} \, ( \, e^{- \tau  \hat H}
 \, e^{- \tau  \hat H}
\cdots
 e^{- \tau  \hat H}  \, )
\label{eq:pathintegral}
\\
& \approx
{\rm Tr} \, ( \, e^{-\tau  \hat K}
 \, e^{-\tau  \hat V}
\, e^{-\tau  \hat K} \, e^{-\tau  \hat V}
\cdots
\, e^{-\tau  \hat K} \, e^{-\tau  \hat V} \, )
\nonumber
\end{align}
Here the exponential of the full Hamiltonian $\hat H = \hat K + \hat V$ is approximated by the product of the exponentials of $\hat K$ and $\hat V$, a well-controlled procedure which can be made arbitrarily accurate by taking $\tau \rightarrow 0$.

The purpose of this procedure is the isolation of the exponential of the interaction term $\hat V$, which can then be rewritten using a Hubbard-Stratonovich(HS) transformation:
\begin{align}
e^{-\tau U
(n_{i \uparrow} - \frac12)
(n_{i \downarrow} - \frac12) }
& = \frac12 e^{-U \tau/4} \sum_{{\cal X}=\pm 1} e^{\lambda
\, {\cal X} \,
(n_{i \uparrow} - n_{i \downarrow} ) },
\label{eq:dhs}
\end{align}
where ${\rm cosh} \, \lambda = e^{U \tau/2}$.  Because one needs to
transform the interaction term on every spatial site $i=1,2,\cdots N$
and also for each of the $l=1,2,\cdots L$ exponentials of $\tau \hat V$
in Eq.~(\ref{eq:pathintegral}), there are a total of $NL$ HS variables
${\cal X}(i,l)$.

In Eq.~(\ref{eq:dhs}), we have employed the discrete HS transformation
introduced by Hirsch,~\cite{hirsch85} but one could also
use a continuous variable ${\cal X}$ and a Gaussian integral,
\begin{align}
e^{-\tau U
\left(n_{i \uparrow} - \frac12\right)
\left(n_{i \downarrow} - \frac12\right) }
&= \frac{e^{-U\tau/4}}{\sqrt{\pi}}
\int d{\cal X}\,
e^{-{\cal X}^2 + 2 \gamma {\cal X}
\left(n_{i \uparrow} - n_{i \downarrow} \right) }
\label{eq:chs}
\end{align}
with $\gamma = \sqrt{U\tau/2}$. There are some differences in the efficiency of the exploration of phase space between the discrete and continuous cases.~\cite{buendia86}

Once the HS transformation is introduced, all the exponentials in the trace of Eq.~(\ref{eq:pathintegral}) are quadratic in the fermion operators, so, the identity in Eq.~(\ref{eq:quadformidentity}) can be used to perform the trace over the Hilbert space analytically.  The sum over the HS configurations ${\cal X}(i,l)$ is performed stochastically using Monte Carlo techniques.  The corresponding Boltzmann weight takes the form of the product of two determinants (one for each spin specie) of matrices $M_\sigma({\cal X})$ of dimension $N$. As this determinant product may be negative for some HS configurations, the sampling is done using the absolute values of the determinant product, and measured expectation values are adjusted accordingly.

The average sign $\langle S \rangle$ is then defined to be the ratio of the integral of the product of up and down spin determinants, to the integral of the absolute value of the product. An analogous definition holds for the average sign $\left< S_{\sigma}\right>$ of the individual determinants:
\begin{align}
\langle S \rangle &=
\frac
{
\sum_{\cal X} \,\,
{\rm det} M_\uparrow({\cal X}) \,
{\rm det} M_\downarrow({\cal X})
}
{
\sum_{\cal X}  \,\,
| \, {\rm det} M_\uparrow({\cal X}) \,
{\rm det} M_\downarrow({\cal X}) \, |
}
\nonumber \\
\langle S_\sigma \rangle &=
\frac
{
\sum_{\cal X}  \,\,
{\rm det} M_\sigma({\cal X})
}
{
\sum_{\cal X}  \,\,
| \, {\rm det} M_\sigma({\cal X})|
}.
\label{eq:signdefined}
\end{align}
In the case we consider here, with no external magnetic field, by symmetry
$\left< S_{\uparrow} \right> = \left< S_{\downarrow} \right>$.

As a practical matter, these quantities are obtained by generating
configurations with the (non-negative) weight
$ | {\rm det} M_\uparrow({\cal X})
{\rm det} M_\downarrow({\cal X}) |$  and measuring the
ratios
$ {\rm det} M_\uparrow({\cal X})
\, {\rm det} M_\downarrow({\cal X})  \, /  \,
| \, {\rm det} M_\uparrow({\cal X})
\, {\rm det} M_\downarrow({\cal X}) \, |$
and
$ {\rm det} M_\sigma({\cal X})
 \, /  \,
\, | \, {\rm det} M_\sigma({\cal X}) \, |$
for each configuration ${\cal X}$. In Eqs.~(\ref{eq:dhs}) and (\ref{eq:chs}), we have coupled the HS variable to the $z$ component of fermionic spin, $n_{i\uparrow}- n_{i\downarrow}$.  It is also possible to write transformations which involve the $xy$ components of a spin, $c_{i\uparrow}^{\dagger} c_{i\downarrow}^{\phantom{\dagger}}$, or even local pairing operators, $c_{i\uparrow}^{\dagger} c_{i\downarrow}^{\dagger}$.  These in general worsen the sign problem.~\cite{tremblay92,batrouni90} In the attractive Hubbard Hamiltonian, the HS variable couples to the charge $n_{i\uparrow}+ n_{i\downarrow}$ on site  $i$.  This makes the matrices $A_i$ identical for up and down fermions, so that their determinants are also identical, and thus, there is no sign problem.  If a charge decoupling is used for the repulsive model, the HS transformation would involve complex numbers, and the determinants would be complex as well, leading to an even more challenging `phase problem'.

We note that there are many details omitted in this brief description, including methods to stabilize the product of the $B_i$ matrices so that round-off errors do not accumulate, the precise Monte Carlo update procedure (how many variables are altered in each step), more rapid procedures for obtaining the ratio of new to old determinants after a HS variable is updated, how to evaluate non-equal time observables, analytic continuation to obtain dynamic behavior, and so forth.  The reader is referred to Refs.~[\onlinecite{sorella89,white89,alvarez08,nukala09,gull11,tomas12,chang13,assaad02,jarrell96}] for more complete discussions.

\subsection{Particle-Hole Symmetry}

On a bipartite lattice, and at $\mu=0$, the Hamiltonian Eq.~(\ref{eq:ham}) is PHS.  That is, the Hamiltonian is invariant under the transformation $c_{i\sigma}^{\dagger} \rightarrow (-1)^i c_{i\sigma}^{\phantom{\dagger}}$. Here $(-1)^i = +1(-1)$ on the {\cal A}({\cal B}) sublattice. This symmetry is present even if $t$ and $U$ vary spatially across the lattice.  As one physical consequence, the density $\rho_i = \langle \, n_{i\uparrow} + n_{i \downarrow} \, \rangle = 1$ (half-filling) for all values of $t,U$ and $T$.  Correlation functions at density $\rho$ and $2-\rho$ have the same values, or are trivially related.  Particle-hole transformations involving only one spin specie can also be used to relate the attractive and repulsive Hubbard Hamiltonians in this limit.

PHS has profound implications for DQMC. When it is present, the determinants of the up and down matrices $M_\sigma$ have the same sign.  Thus, although they individually can go negative, their product is always positive. As a consequence, low temperature physics can be accessed at half-filling.  This fact enabled DQMC to establish rigorously~\cite{hirsch85,hirsch89} that the single band, square lattice Hubbard model has long-range Ne\'el order at $T=0$, as opposed to a disordered (resonating valence bond) ground state.

The sign problem can be absent in some other types of Hamiltonians with a similar symmetry requirement, for example in a model with an interaction which takes the form of the square of near-neighbor hopping, \cite{assaad96} for a low energy theory of the onset of antiferromagnetism,~\cite{berg12} and in spin polarized Fermi systems.~\cite{huffman14} Indeed, the number of special situations where the sign problem is absent is rapidly growing, including, for example, in quasi-1D condensed matter models of ferromagnetism,~\cite{xu14} and via the `fermion bag' approach, in lattice gauge theory. \cite{chandrasekharan10,chandrasekharan12} The sign problem is also absent in Hubbard models with a larger number of spin components,~\cite{zhou14} and, very interestingly, in a class of spinless fermion models\cite{li14}, a unique situation where positivity is not dependent on having an even number of fermionic species.

\section{Sign Problem Datasets}

Our goal in this section is to present a unified and easily comparable collection of data for the sign problem for different lattice geometries, including hypercubic lattices in dimensions $d=1,2,3$; other bipartite structures like the honeycomb, Lieb and 1/5-depleted square lattices; and finally two non-bipartite lattices: triangular and Kagome.  For each case we will exhibit the average sign for a range of temperatures $T$, interaction strengths $U$, lattice sizes, and densities. 

We focus on the product of the signs of the up and down determinants, since that is what is relevant for extracting physics from the DQMC simulation.  However, in Sec.~IVB we will present a brief analysis of the individual spin components.  Among other things, we observe that, except at PHS points, the signs of the up and down determinants tend to be rather uncorrelated, so that the signs of the individual components lend no further information, and can be qualitatively inferred from the square root of the total sign.

Some of these geometries have unique features in their non-interacting densities of states (DOS).  The square lattice possesses a van-Hove DOS singularity at $\rho=1$.  In contrast, the honeycomb lattice has a DOS which vanishes linearly there.  The Lieb lattice has a flat energy band between two dispersing ones, while the flat band in the Kagome lattice can be chosen to be either the lowest or highest set of energy levels, depending on the sign of $t$.  One of our goals is to examine how such features might affect the fermion sign, an issue to which we will return in the conclusion.

\subsection{Hypercubic Lattices}

In this subsection we present data for hypercubic lattices; linear chains, ladders, the square lattice, and the cubic lattice. In all cases, we use periodic boundary conditions except for the rungs of the ladder geometry.

\begin{figure}[t]
\epsfig{figure=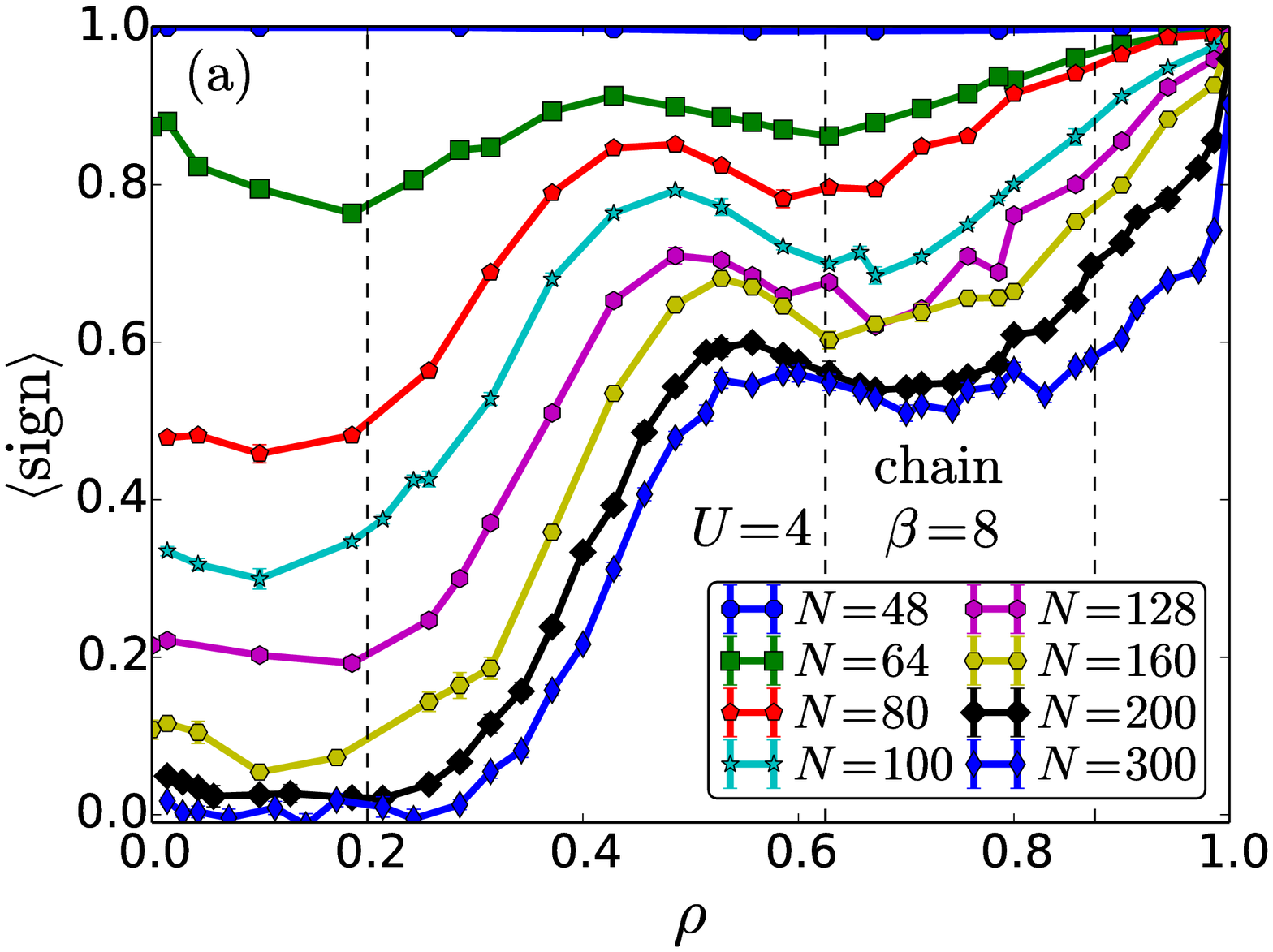, width=3.3in,angle=-0,clip}
\epsfig{figure=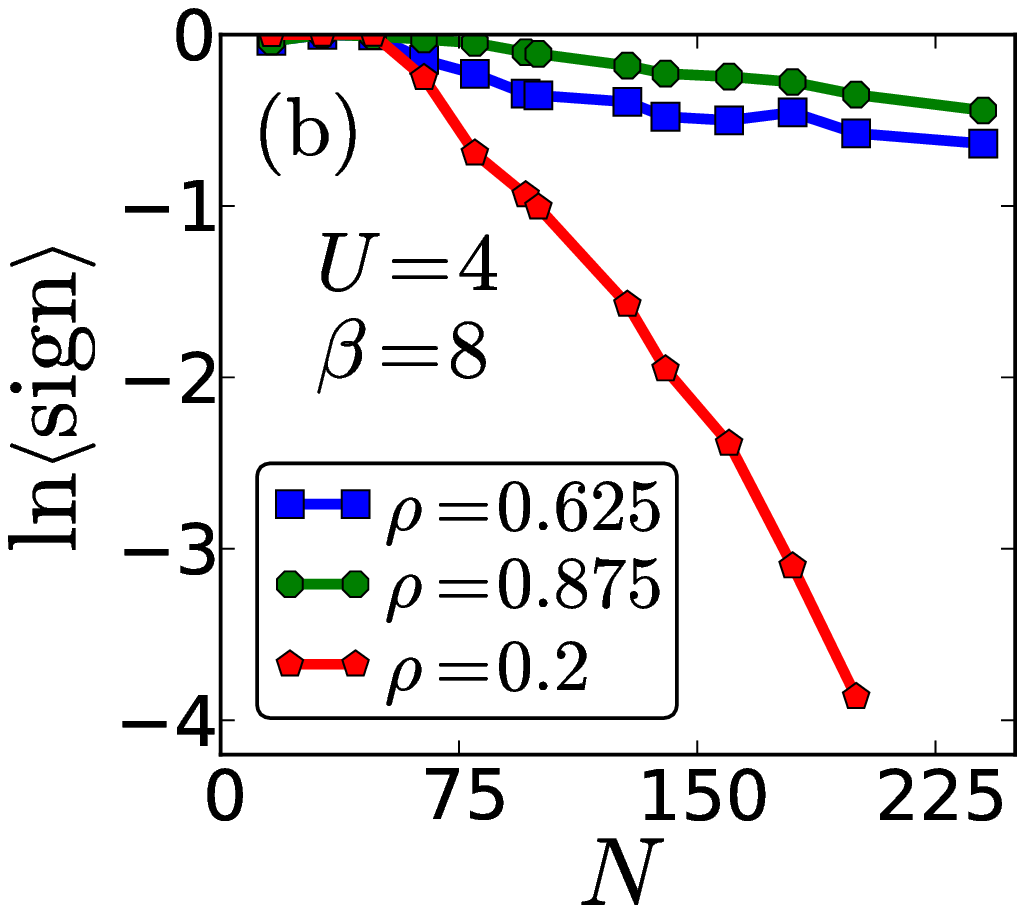, width=1.6in,angle=-0,clip}
\epsfig{figure=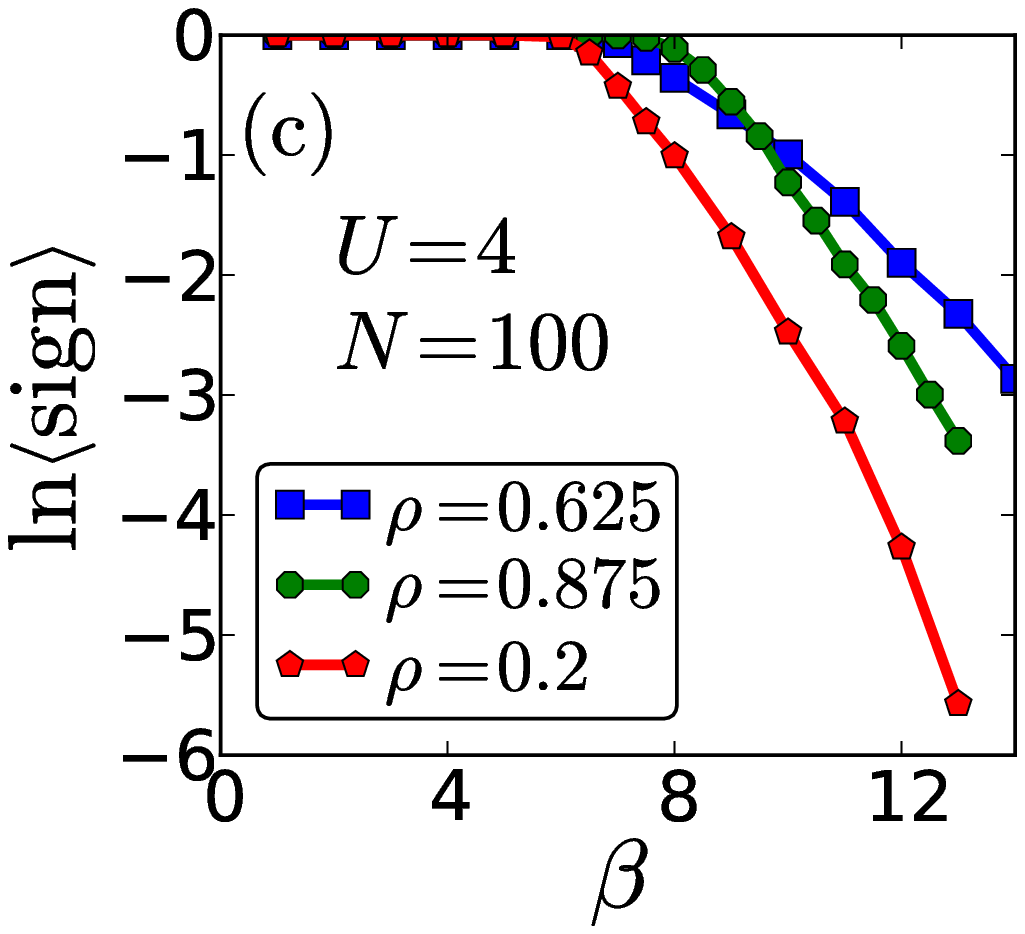, width=1.6in,angle=-0,clip}
\caption{(Color online){ Sign problem for chains of different lengths at
$U=4$ (which equals the bandwidth in this geometry). (a) The dependence
of $\left< S \right>$  on the density. (b) $\ln \left< S \right>$ vs the
system size, $N$, and (c) vs $\beta$ at several fixed densities $\rho$.
The densities used in panels (b) and (c) are indicated by vertical
dashed lines in panel (a). The errorbars in this figure and in the
remaining figures in this paper can be
inferred from the scatter in the data, and hence, may not have been shown.
Axis label $\left< {\rm sign} \right>$}
is referred to as $\left< S \right>$ in text.}
\label{fig:chainU4beta8}
\end{figure}

\begin{figure}[t]
\epsfig{figure=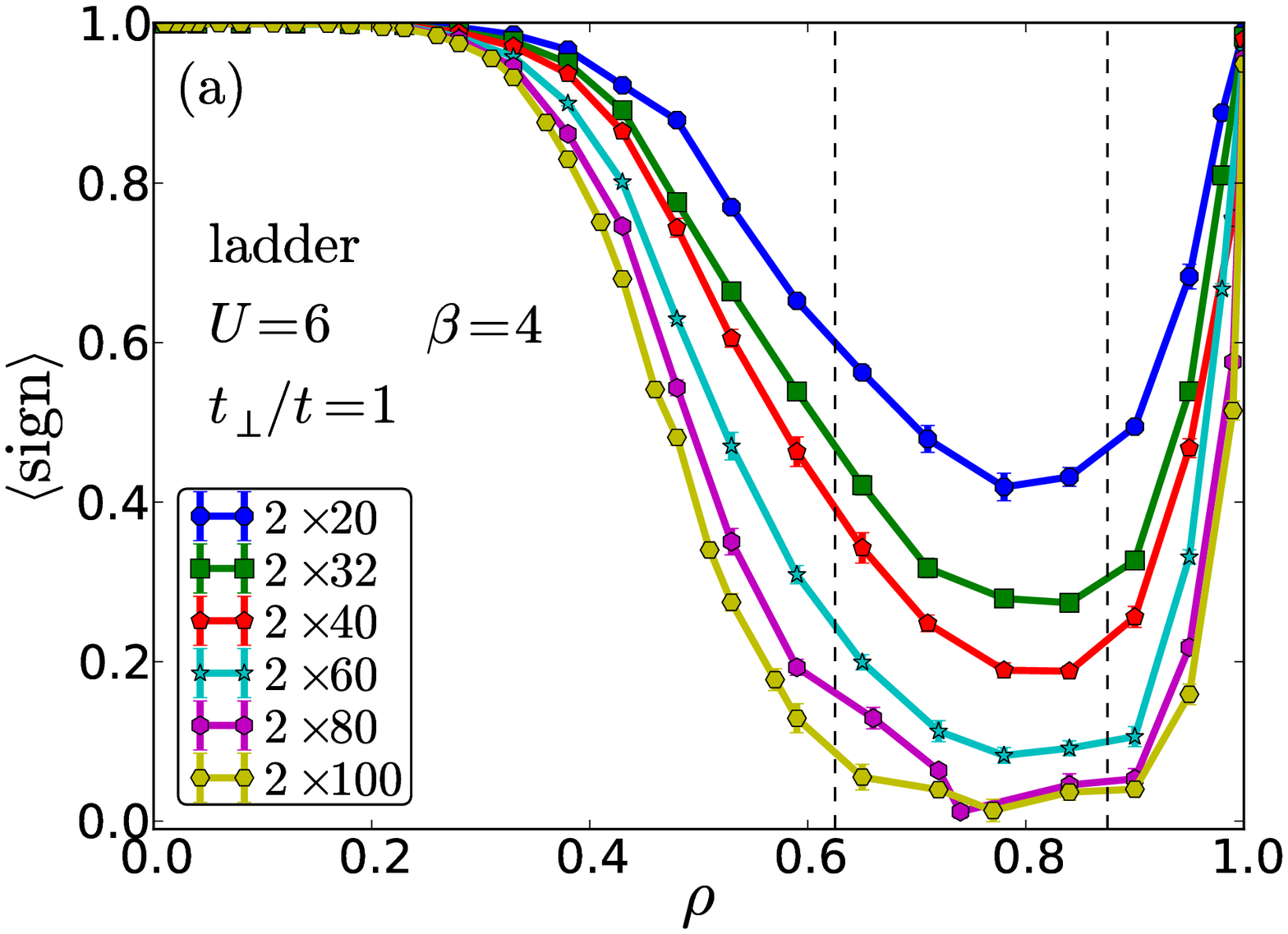, width=3.3in,angle=-0,clip}
\epsfig{figure=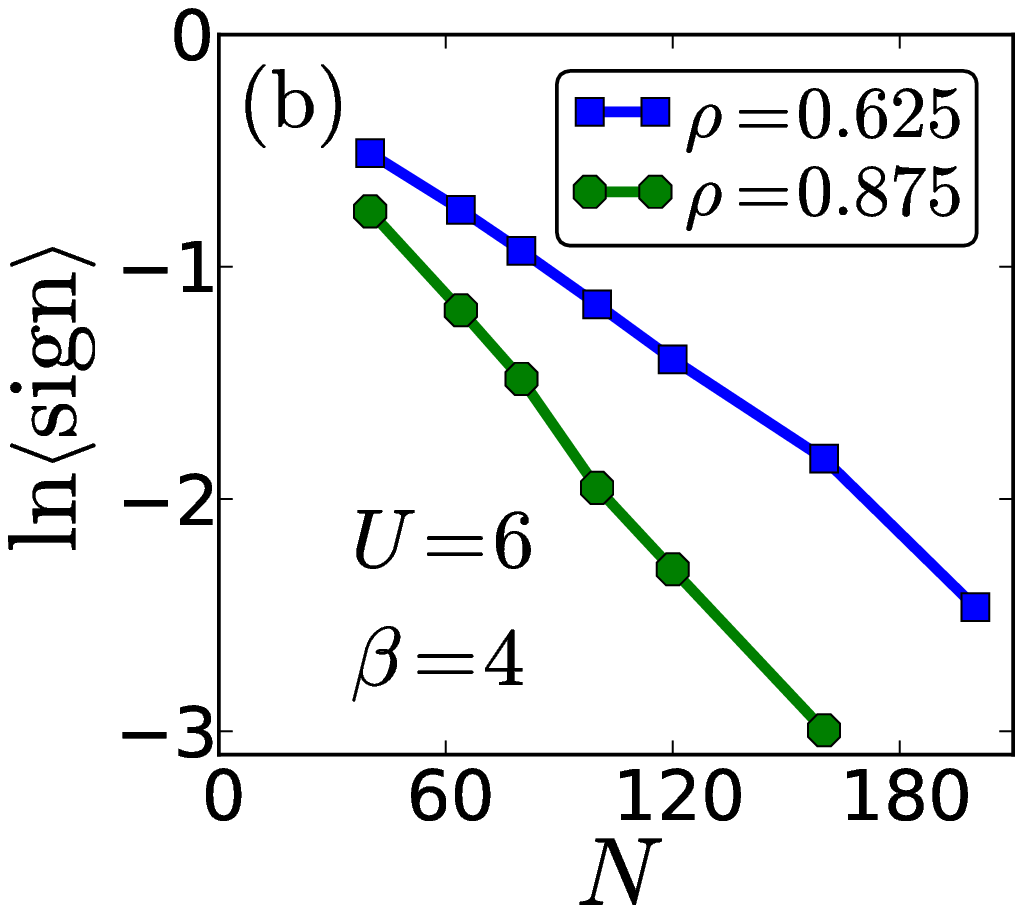, width=1.6in,angle=-0,clip}
\epsfig{figure=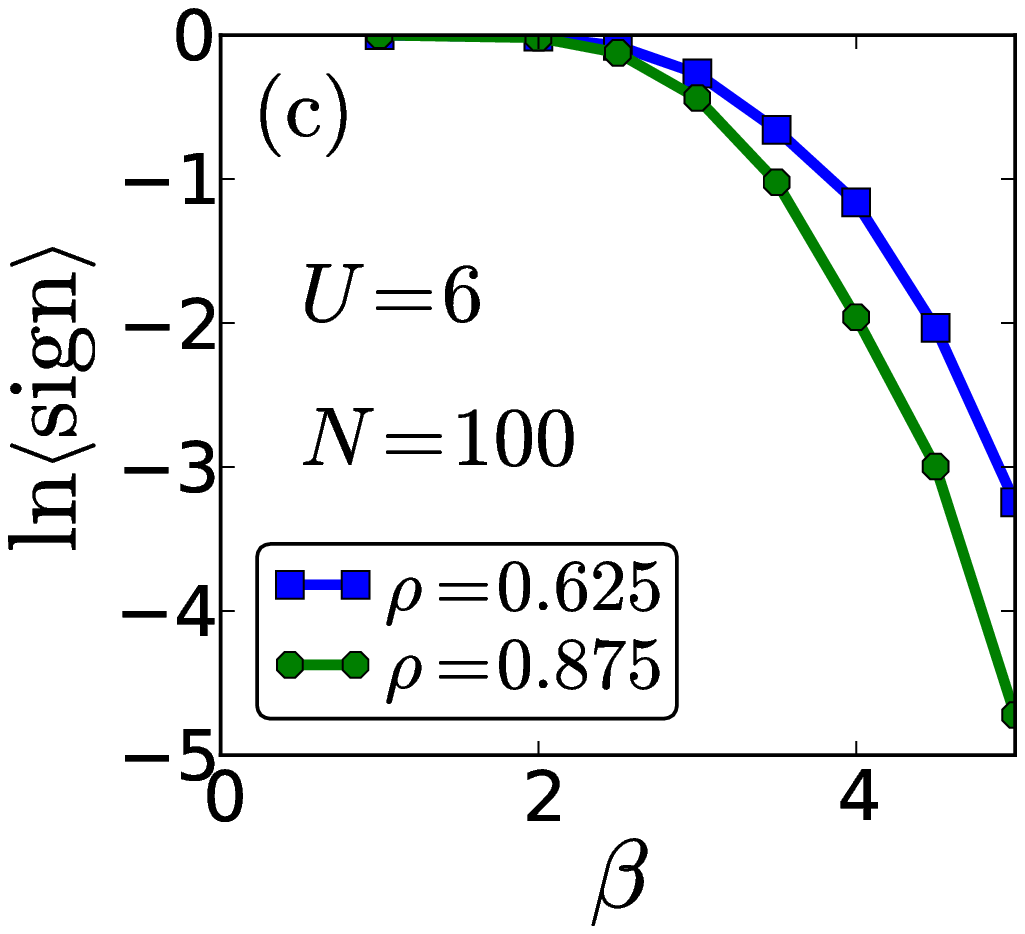, width=1.6in,angle=-0,clip}
\caption{(Color online) Same as Fig.~\ref{fig:chainU4beta8} but for the ladder
geometry with the rung hopping equal to the hopping along the chains and different
values of the interaction strength and $\beta$, which are noted in the panels.}
\label{fig:ladderU6beta4}
\end{figure}

\begin{figure}[t]
\epsfig{figure=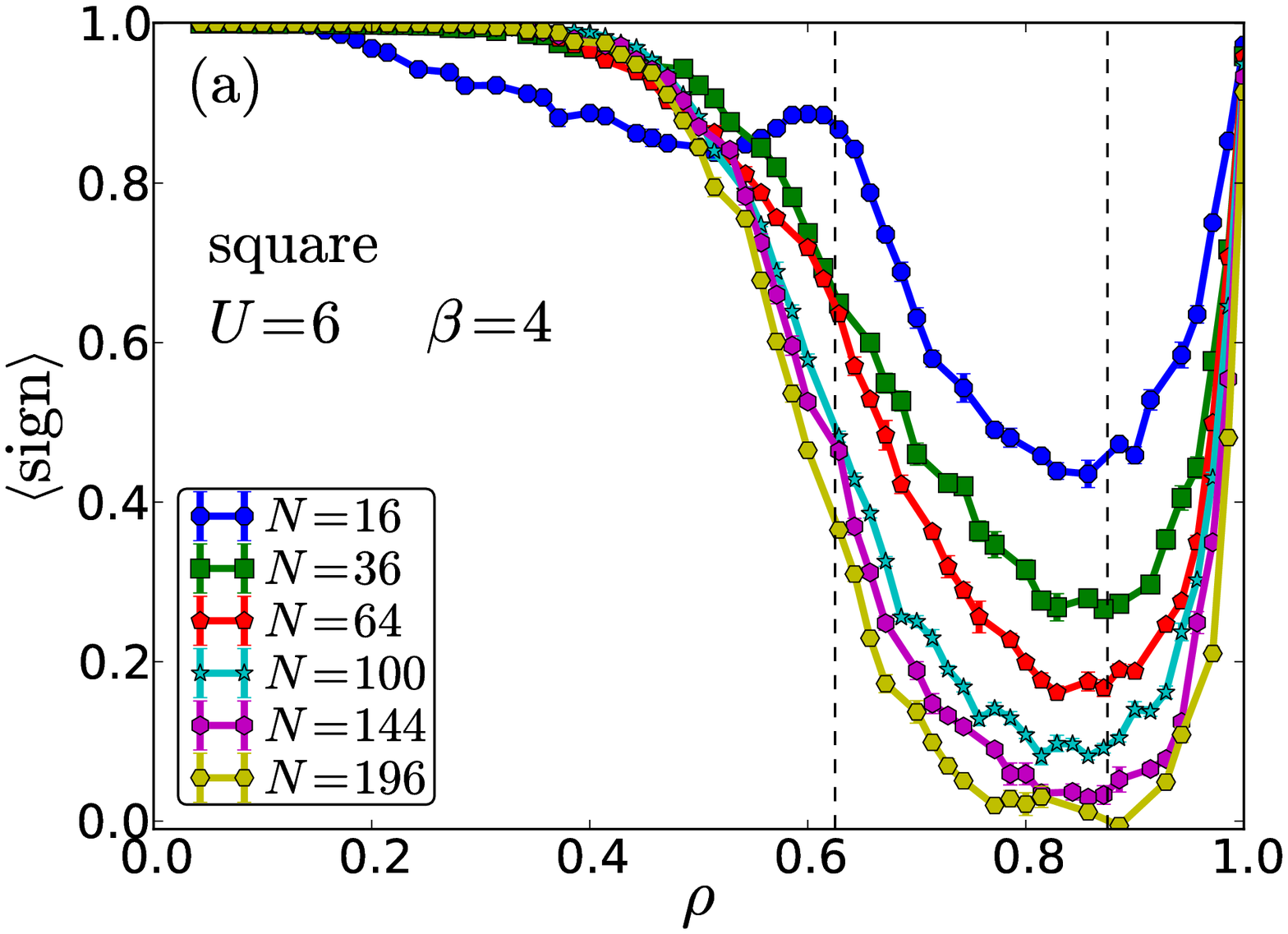, width=3.3in,angle=-0,clip}
\epsfig{figure=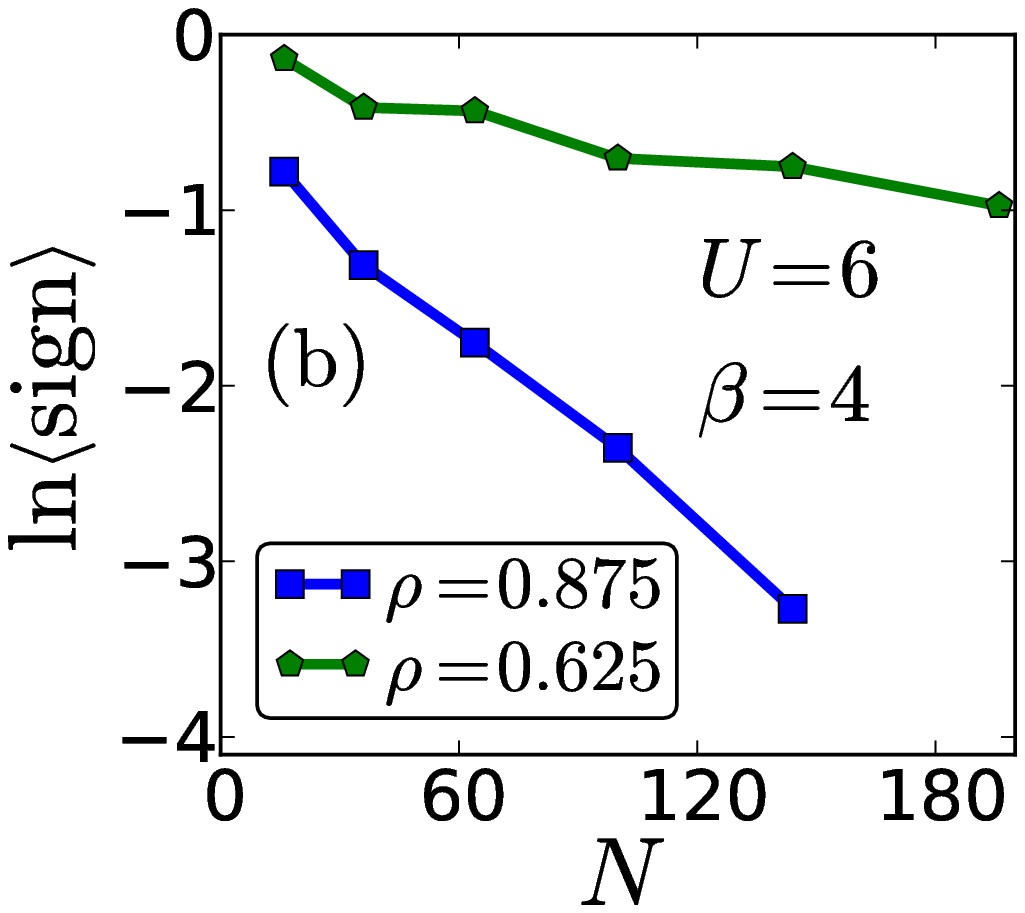, width=1.7in,angle=-0,clip}
\epsfig{figure=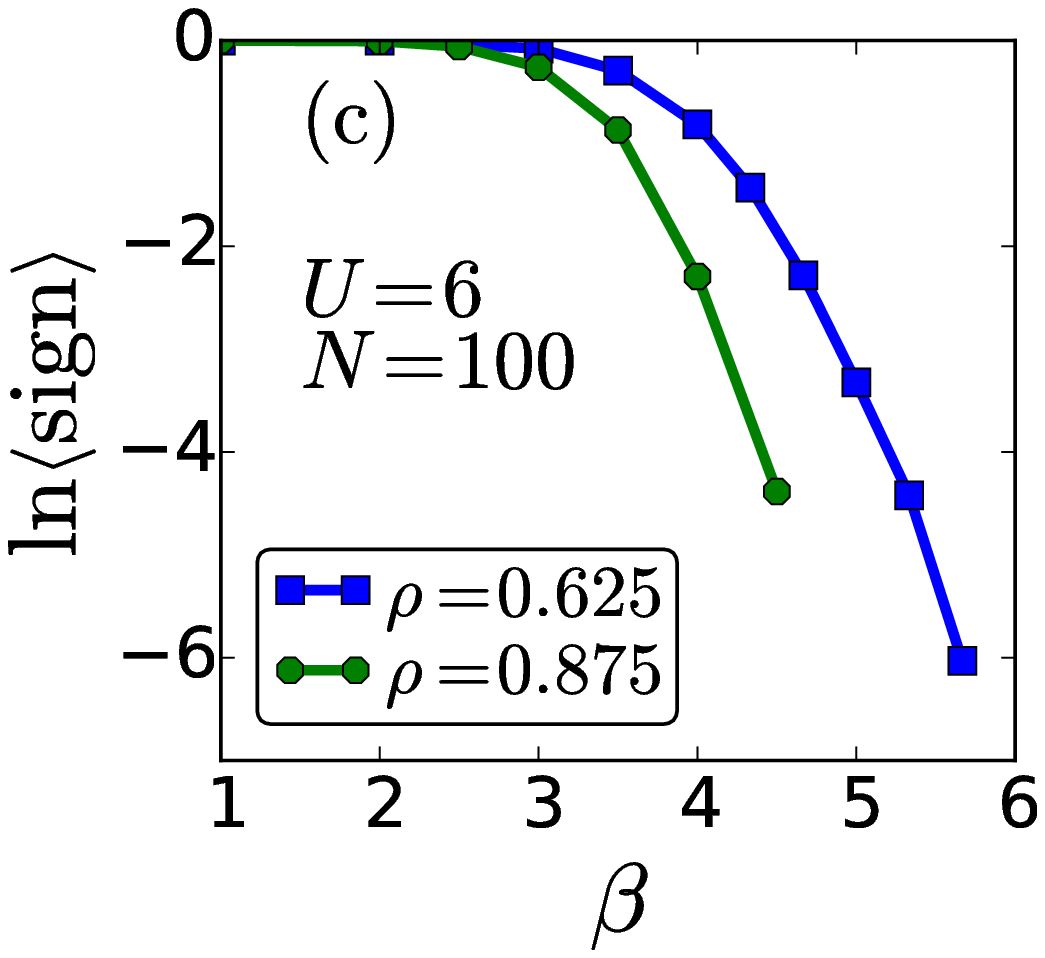, width=1.6in,angle=-0,clip}
\caption{(Color online) Same as Fig.~\ref{fig:chainU4beta8} but for the square geometry
and with different parameters, which are noted in the panels.}
\label{fig:squareU6beta4}
\end{figure}

Although the sign does not cause a problem for Hubbard world line methods\cite{hirsch82,assaad08} in one dimension, DQMC does have a sign problem in this case\cite{footnote2}. In Fig.~\ref{fig:chainU4beta8}, the average fermion sign, $\langle S \rangle$ (also denoted $\langle \textrm{sign} \rangle$), for the chain geometry is shown for fixed $U=4$ and $\beta=8$. Figure~\ref{fig:chainU4beta8}(a) shows the doping dependence of the sign, which has a non-trivial structure. Most notably, it shows a peak around $\rho=0.5$.  One may wonder if such a local maximum arises due to remnants of the `shell' effect. That is, at $U=0$ the $k$ space grid is sufficiently coarse such that $\rho(\mu$) shows distinct plateaus where $\langle S \rangle$ tends to be closer to one. This phenomenon is well known, for example, on square lattices that are not too large (e.g.,~$4\times 4$). However, as will be seen in Fig.~\ref{fig:squareU6beta4}, this occurs only on small lattices, and is unlikely the origin of the maximum at quarter filling here where the $k$ space grid is much more refined.  Another interesting feature is that $\langle S \rangle$ remains small at low density.  This is, again, rather different from what happens on a square lattice where $\langle S \rangle \rightarrow 1$ as $\rho \rightarrow 0$ (Fig.~\ref{fig:squareU6beta4}) or even ladders (Fig.~\ref{fig:ladderU6beta4}).  The low value of $\langle S \rangle$ as $\rho \rightarrow  0$ does not appear to be connected to the divergence of the density of states at the bottom of the band since the same divergence occurs in the ladder geometry where $\langle S \rangle$ recovers to one as $\rho \rightarrow 0$.

Figures~\ref{fig:chainU4beta8}(b) and \ref{fig:chainU4beta8}(c) show the scaling of the average sign with spatial lattice size $N$ and inverse temperature $\beta$ respectively.  After plateaux at small $N$ and $\beta$, where $\langle S \rangle=1$, the average sign decreases in a manner which is largely consistent with exponential. ${\rm ln}\langle S \rangle$ is not perfectly linear in $\beta$ or $N$, but exhibits some downward curvature, which we believe indicates the scaling regime has not yet been fully attained.  We will remark on this more fully later in this section. As we shall see, this exponential decrease is also the case in other geometries, although the decrease with $N$ once one exits the plateau is in general less abrupt than with $\beta$.  In Fig.~\ref{fig:chainU4beta8}(b) at $\rho=0.625, 0.875$, the average sign, despite its exponential decay, remains quite manageable out to $N \gtrsim 200$.  Even at $\rho=0.2$ the spatial size must be  tripled  from $N\sim 70$ before ln$\langle S \rangle \sim -4$.  In Fig.~\ref{fig:chainU4beta8}(c), on the other hand, the decay to ln$\langle S \rangle \sim -4$ takes place after only a 50\% increase in $\beta$ (from $\beta=8$ to $\beta=12$).

Next, we turn to ladder geometries, which are natural extensions of chains, before studying the square lattice. Ladders are of interest for several reasons.  First, they have been extensively studied by DMRG~\cite{noack94} as a stepping stone to 2D. Second, by changing the ratio $t_\perp/t$ of the rung hopping to the hopping along the chains, one can access $U=0$ states that are metallic or band-insulating at half-filling.  The effect of these phase changes on the sign problem  for $U\ne 0$ is one goal of the data presented here and in Sec.~IVB.

The results for the average sign on different ladders are plotted in Fig.~\ref{fig:ladderU6beta4}. The rung hopping is set to $t_\perp = 1$, so that the $U=0$ band structure is metallic. As with Fig.~\ref{fig:chainU4beta8}, the top panel shows the density dependence for different system sizes at fixed $U,\beta$ while the bottom two panels show that the decay is consistent with exponential in $N$ and $\beta$.
(See remarks to follow on challenges in capturing fully linear behavior of ${\rm ln}\langle S \rangle(\beta)$.) $\langle S \rangle $ has a minimum at filling $\rho \approx 0.8$, similar to what is known to occur also for a square lattice.  (See Fig.~\ref{fig:squareU6beta4}.) For $t_{\perp}/t > 2$ the noninteracting system is a band insulator (BI). Because the BI occurs at the particle-hole symmetric density $\rho=1$, $\langle S \rangle$ is pinned at unity. Therefore, for this case, we will examine the signs of  the determinants of the {\it individual} spin matrices $\langle S_\uparrow \rangle =\langle S_\downarrow \rangle$ in Sec.~IVB.


The square lattice is the most well-studied Hubbard model geometry, owing to its possible relevance as a simple model of cuprate superconductivity and $d$-wave pairing driven by antiferromagnetic fluctuations.~\cite{scalapino94} The total sign for the square lattice at $U=6, \beta=4$ is shown as a function of filling $\rho$ for different lattice sizes in Fig.~\ref{fig:squareU6beta4}(a).  As mentioned earlier, the peak in the $4\times 4$ lattice occurs as five of the 16 allowed $k$ points (corresponding to a density $\rho=10/16=0.625$) fill up prior to half-filling.  This peak is even more evident\cite{white89} at $U=4$ providing further evidence that, in this case, $\langle S \rangle$ is connected to the shell structure at $U=0$, though the connection appears to diminish with larger $U$.  It is possible that the shell structure would appear on larger lattices if lower temperatures were accessible. A rather universal feature is the minimum in $\langle S \rangle$ at $\rho \approx 0.85$, which is shared by the ladder geometry.

The very nature of the sign problem makes it challenging to provide a completely compelling linear plot of ln$\, \langle S \rangle$ as a function of $\beta$.  For the data in Fig.~3c, for example, at $\beta=5.33$ a run with $10^5$ sweeps through the space-time lattice of Hubbard-Stratonovich variables (100 spatial sites and 64 time slices) takes several hours on a work-station, and gives $\langle S \rangle = 0.0095 \pm 0.0012$ (corresponding to ${\rm ln}\, \langle S \rangle = -4.66$.  The slope ${\rm d}\,{\rm ln}\, \langle S \rangle / d\beta \sim -4$, so we can roughly estimate ${\rm ln}\, \langle S \rangle \sim -8.9$ at $\beta=6.33$, and hence $\langle S \rangle \sim 0.00013$.  A measurement of this value to 10\% accuracy would require an error bar of about 0.00001, a factor of $\sim 10^2$ less than the error obtained at $\beta=5.33$.  Since error bars only go down as the square root of the number of measurements, such a run would entail $\sim 10^4$ times as many sweeps, and a cpu time of several months.

\begin{figure}[t]
\epsfig{figure=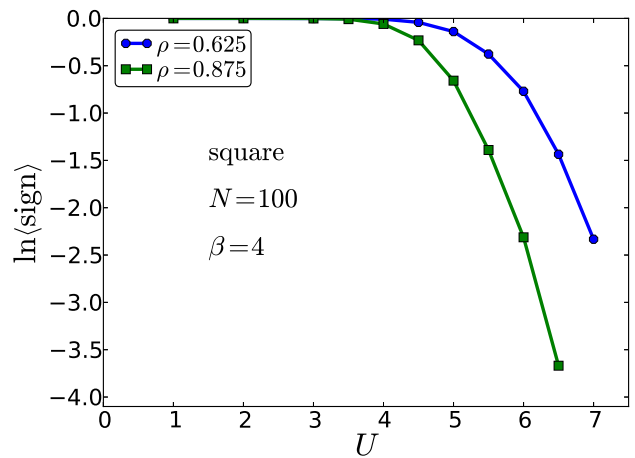, width=3.3in,angle=-0,clip}
\caption{(Color online) The $U$ dependence of the sign problem for the square
lattice.}
\label{fig:square_sign_vs_U}
\end{figure}

The $U$ dependence of the average sign at fixed $\beta=4$ and $N=100$ for a square lattice is shown in Fig.~\ref{fig:square_sign_vs_U} at $\rho=0.625$ and 0.875. The evolution of $\langle S \rangle$ with $U$ is rather similar to that with $\beta$: A plateau at weak coupling where $\langle S \rangle \approx 1$ is followed by an abrupt downturn.  Thus, in practice, once the sign begins deviating from unity there is only a narrow window of stronger couplings where data can be acquired. This is demonstrated in Fig.~\ref{fig:square_sign_vs_U} as $\langle S \rangle$ as a function of $U$ is shown to decrease exponentially once $U\gtrsim 4$.

\begin{figure}[t]
\epsfig{figure=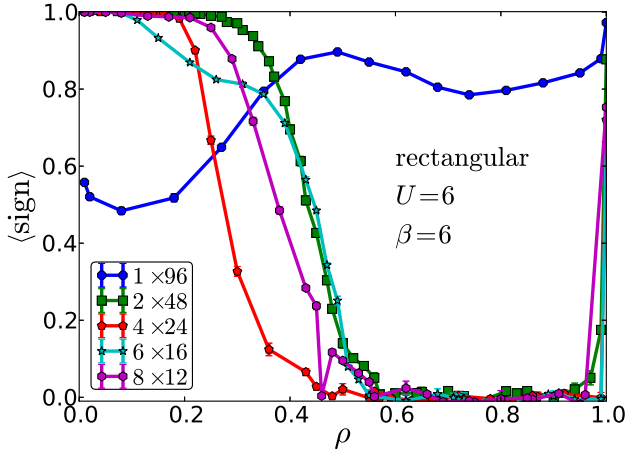, width=3.3in,angle=-0,clip}
\caption{(Color online) The average sign as a function of $\rho$ for the rectangular lattices with the
same lattice size $N=96$ but different lengths of the sides.}
\label{fig:square_96}
\end{figure}

Here, we briefly generalize our ladder results by providing a more complete description of the behavior of $\langle S \rangle$ as the aspect ratio of the lattice goes from one to two dimensions. Fig.~\ref{fig:square_96} shows data at $U=6$ and $\beta=6$ for lattices with a fixed $N=96$ but different aspect ratios. As mentioned earlier, the chain geometry seems to be rather unique. For all other cases, the average sign is close to unity for a range of low densities, and then steps downwards to small values in a region centered at $\rho \sim  0.8$, recovering only at the PHS point $\rho=1$.

\begin{figure}[b]
\epsfig{figure=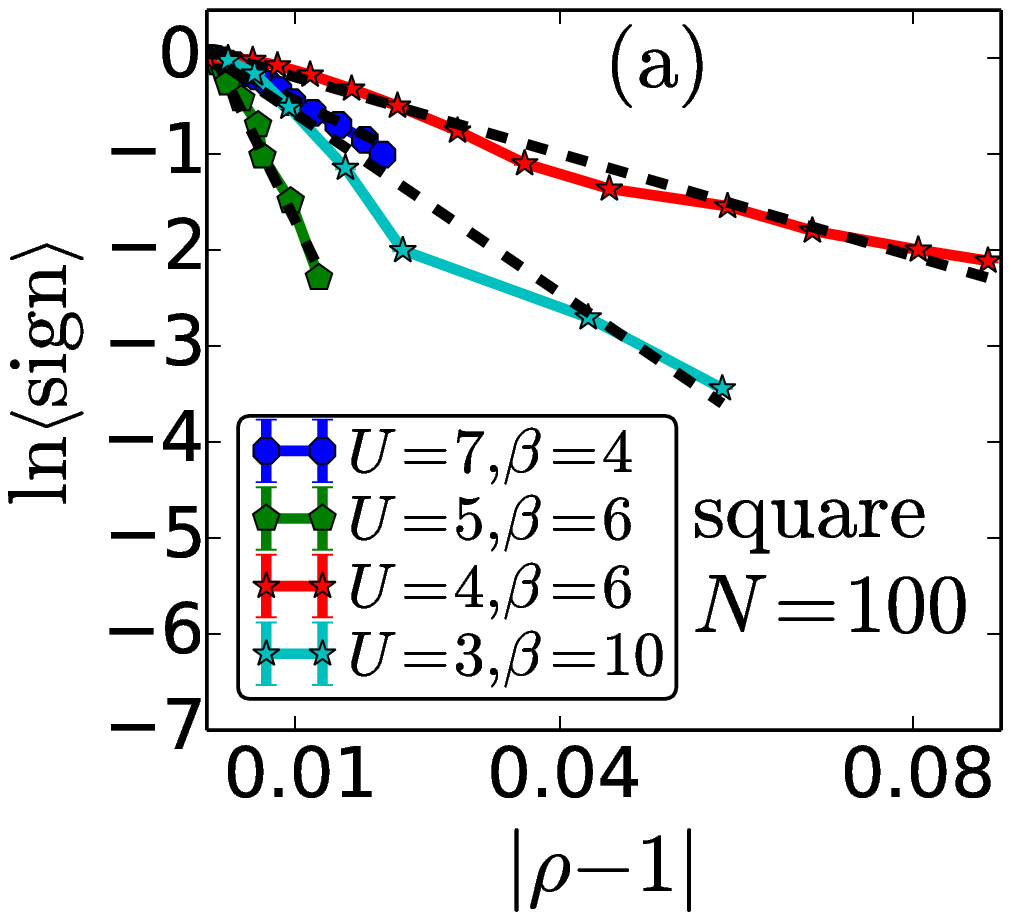, width=1.62in,angle=-0,clip}\hspace{-0.19cm}
\epsfig{figure=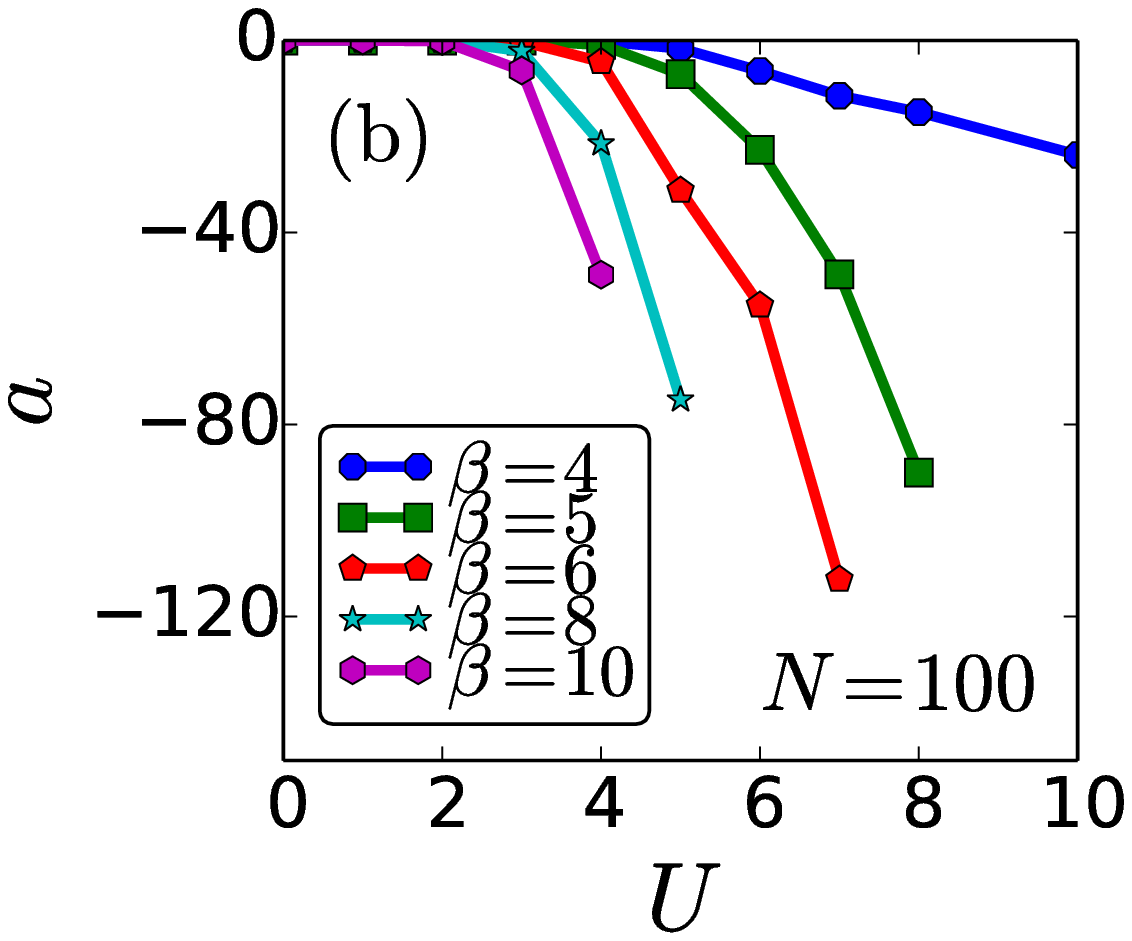, width=1.82in,angle=-0,clip}
\caption{(Color online) (a) By doping the system away from half filling on the square lattice, the
average sign decreases exponentially, i.e., $\langle S \rangle \sim e^{a |\rho - 1|}$.
In (b) the decay constant, $a$, is plotted as a function of $U$ for
different values of $\beta$.  The lattice size is $N=10 \times 10$.}
\label{fig:square_exponents}
\end{figure}

The precipitous nature of the decrease in $\langle S \rangle$ near $\rho=1$, evident in Fig.~\ref{fig:square_96}, is quantified in Fig.~\ref{fig:square_exponents}. Panel \ref{fig:square_exponents}(a) shows the logarithm of the sign versus the doping away from half-filling.  The linear behavior indicates that $\langle S \rangle \sim  e^{a |\rho - 1|}$.  The decay constant $a$ is large and negative. Its $\beta$ and $U$ dependences are given in Fig.~\ref{fig:square_exponents}(b).  To our knowledge, this behavior of $\langle S \rangle$ had not been studied before.  However, scaling forms for physical observables like the compressibility $\kappa=\partial \rho / \partial \mu$ as one exits the Mott phase at $\rho = 1$  have been suggested.~\cite{imada95,assaad96b}  In these theories, $\kappa$ follows a power law $\kappa \propto (1-\rho)^{-\eta}$, so that it diverges just before it vanishes. This also occurs in the boson Hubbard model.~\cite{fisher89,batrouni90b}

\begin{figure}[t]
\epsfig{figure=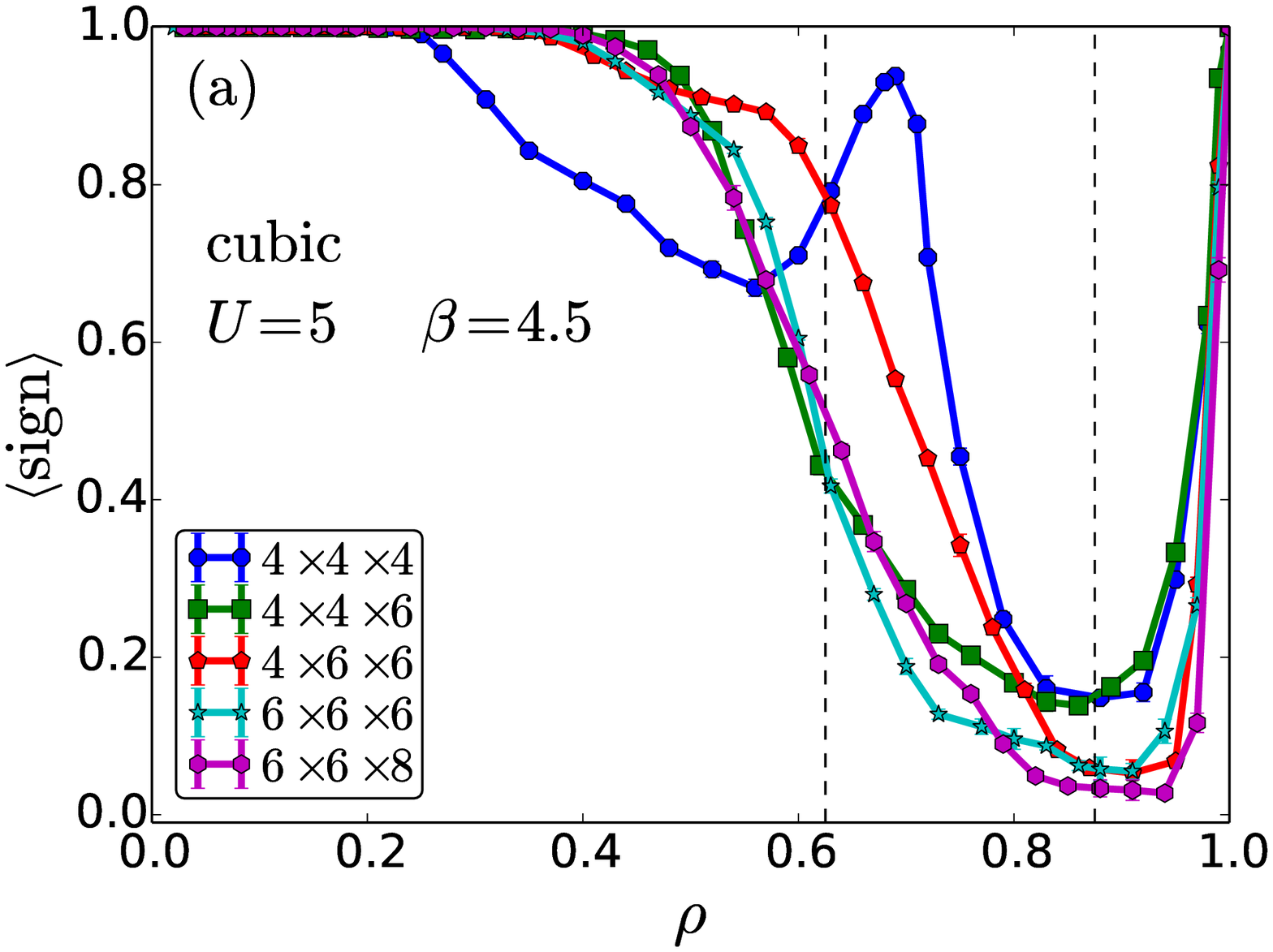, width=3.3in}
\epsfig{figure=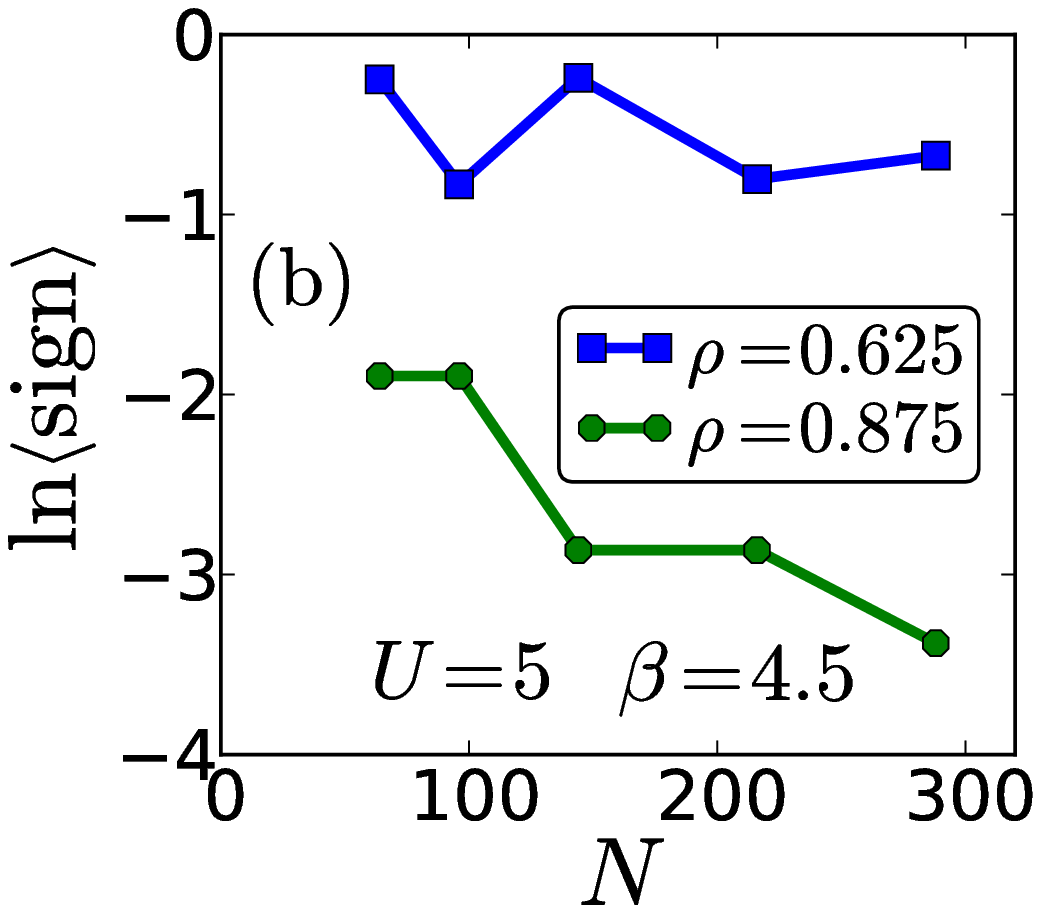, width=1.6in,angle=-0,clip}
\epsfig{figure=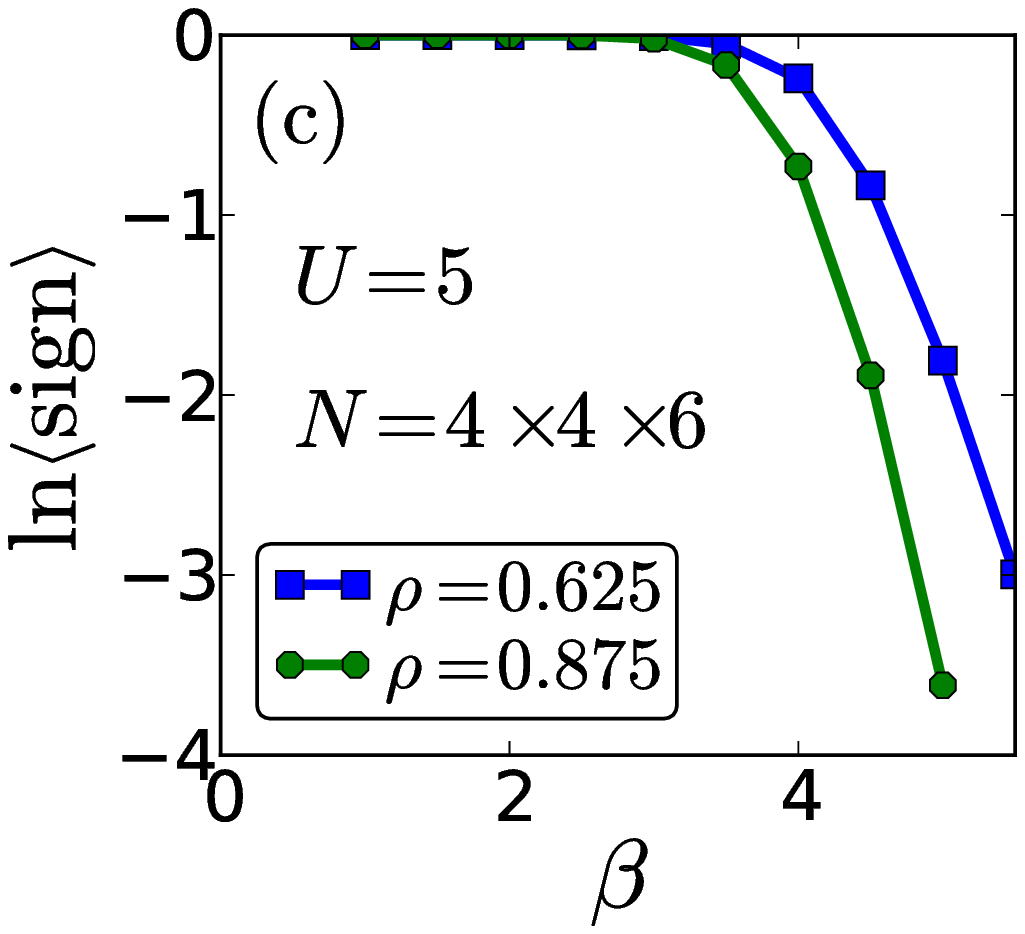, width=1.6in,angle=-0,clip}
\caption{(Color online) Same as Fig.~\ref{fig:chainU4beta8} but for the cubic geometry
and with different parameters, which are noted in the panels.}
\label{fig:cubicU5beta4.5}
\end{figure}

We conclude this section on hypercubic geometries by showing the behavior of $\langle S \rangle$ for a cubic lattice in Fig.~\ref{fig:cubicU5beta4.5}. Because the number of lattice sites grows so rapidly with linear size, we consider cases where $L_x \neq L_y \neq L_z$ (while keeping all linear lengths even to avoid frustration of antiferromagnetic correlations.) The qualitative behavior is almost identical to that of rectangular lattices, with a deep minimum in $\langle S \rangle$ upon doping from half filling, followed by a recovery at $\rho \lesssim 0.6$.  Curves for different sizes almost coincide for these large dopings, indicating a very slow decay with $N$, as seen in Fig.~\ref{fig:cubicU5beta4.5}(b). Indeed, for some densities $\langle S \rangle$ even increases as $N$ increases.  Presumably, this is a transient phenomenon associated with the rather small linear lattice lengths which are accessible in 3D. The decay with $\beta$, Fig.~\ref{fig:cubicU5beta4.5}(c), is, as usual, rapid.

\subsection{Other Bipartite Lattices}

Hypercubic lattices are just one instance of bipartite geometries which are free of the sign problem at half-filling owing to the PHS.  Here we present data for three additional bipartite geometries, all of which are of interest because of their materials applications.


\begin{figure}[t]
\epsfig{figure=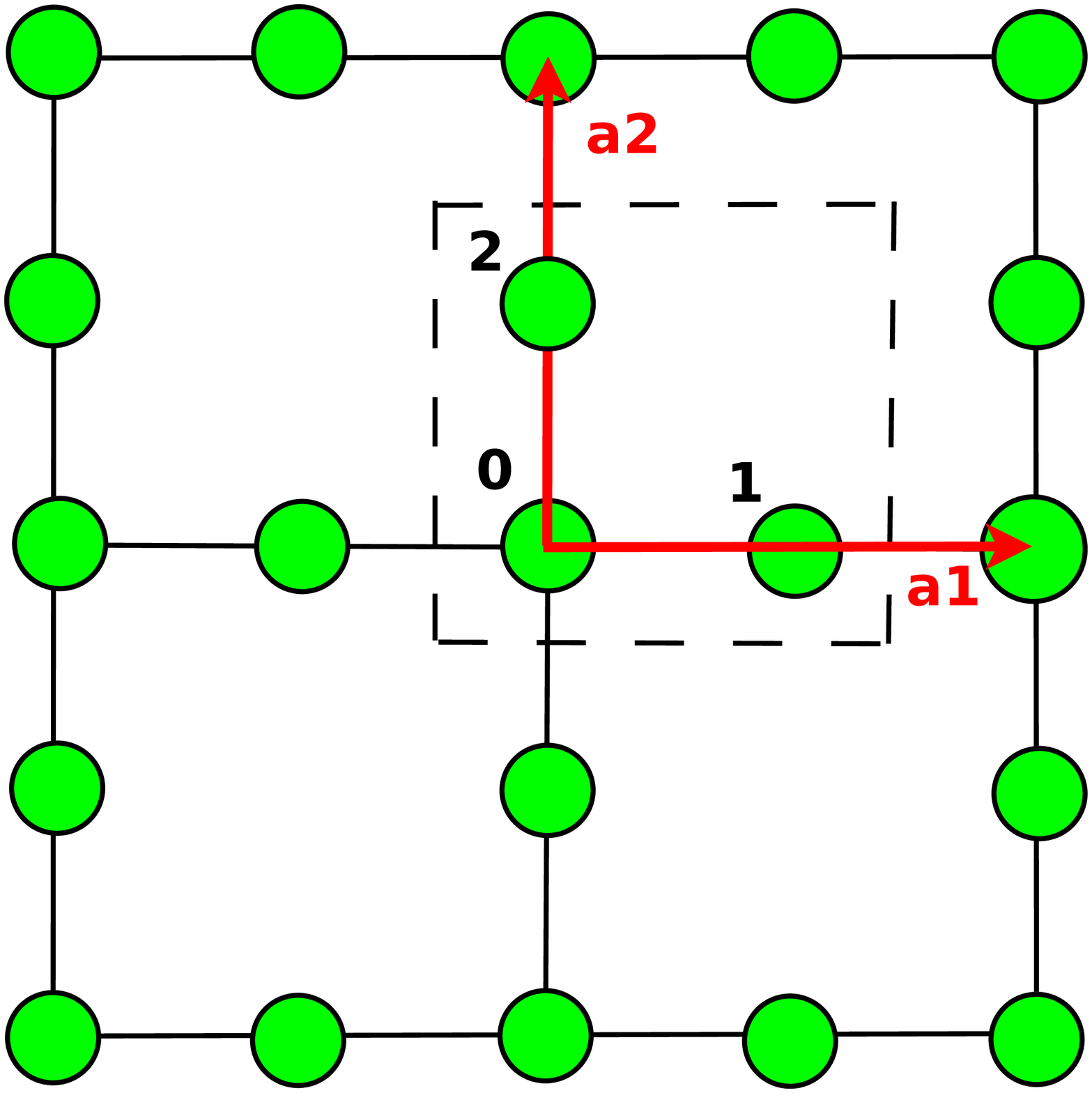, width=1.3in,angle=-0,clip}
\hskip 0.6in
\epsfig{figure=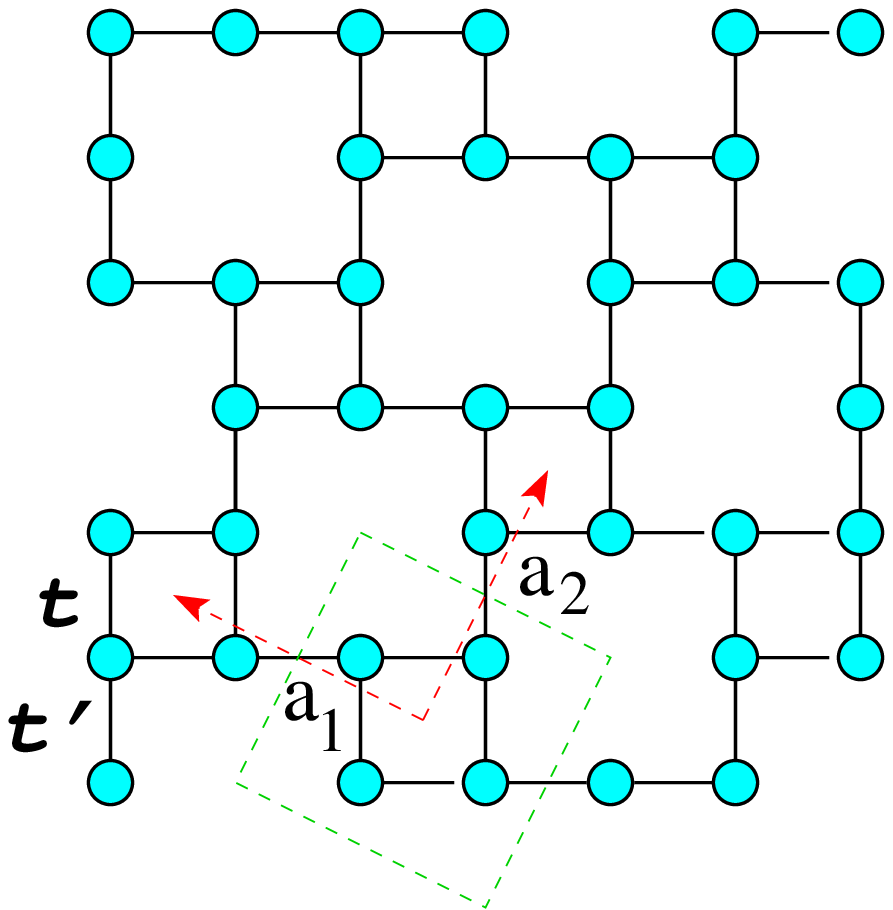, width=1.3in,angle=-0,clip}
\caption{(Color online) The Lieb lattice (left) and the 1/5-depleted square lattice (right).
The latter is taken from Ref.~\onlinecite{khatami14}.
Arrows (dashed squares) show the unit vectors (cells) for each geometry.}
\label{fig:Liebandonefifth}
\end{figure}

\begin{figure}[b]
\epsfig{figure=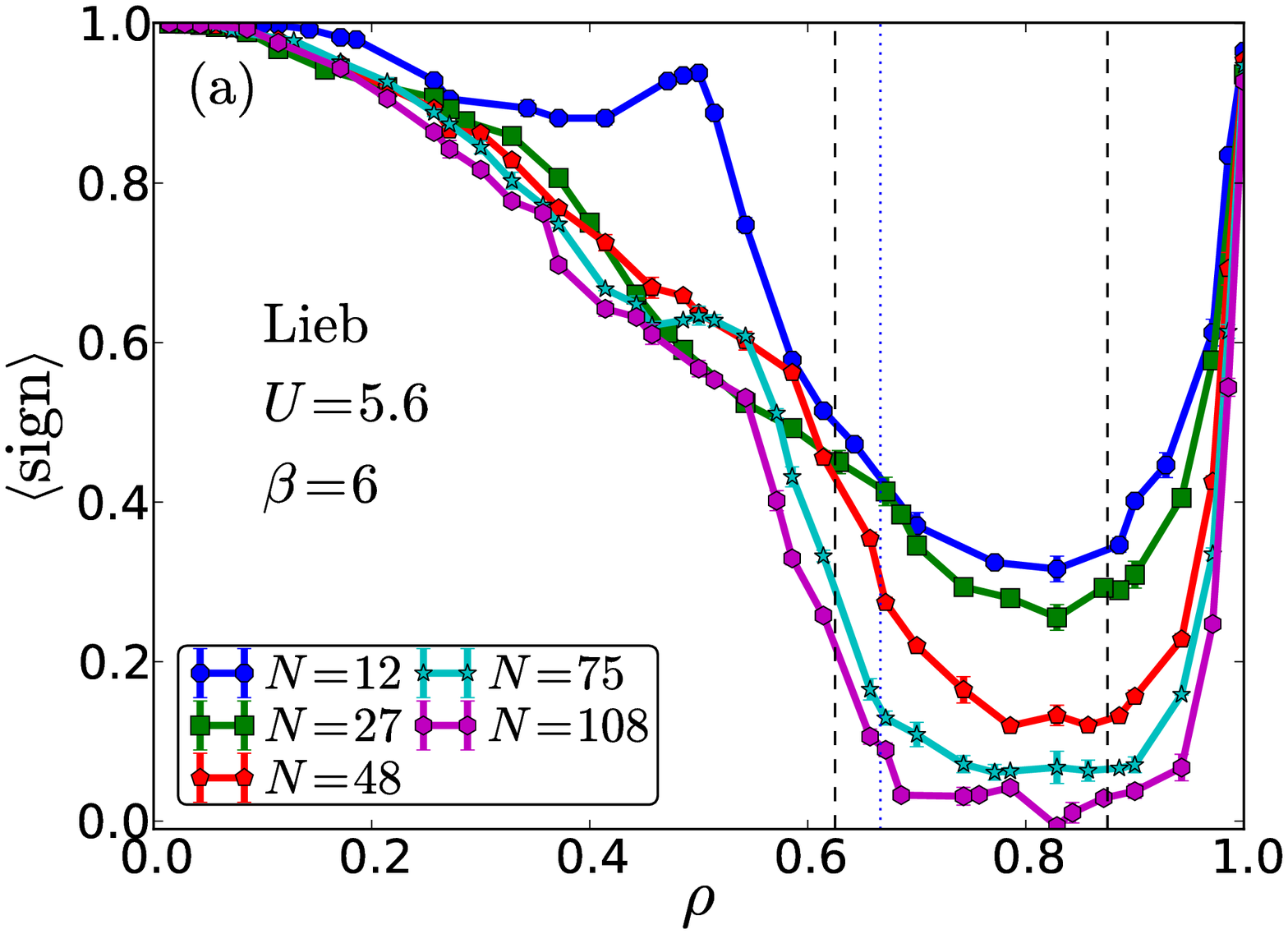, width=3.3in,angle=-0,clip}
\epsfig{figure=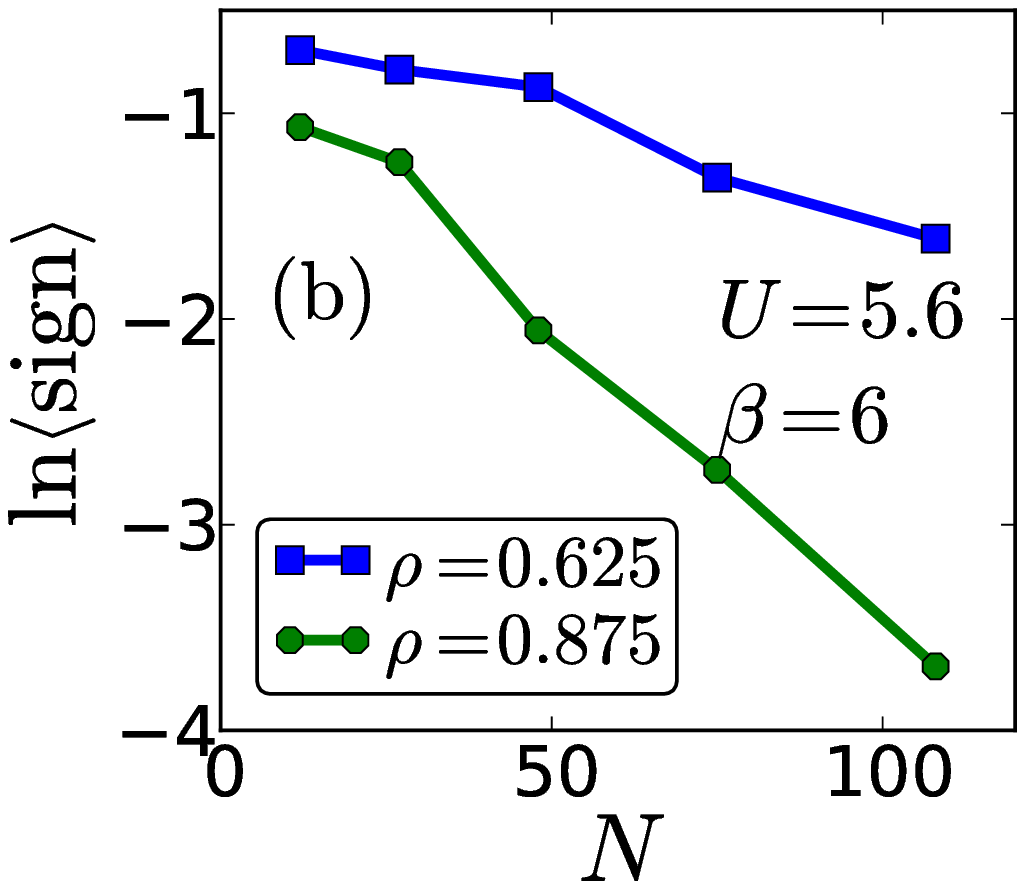, width=1.6in,angle=-0,clip}
\epsfig{figure=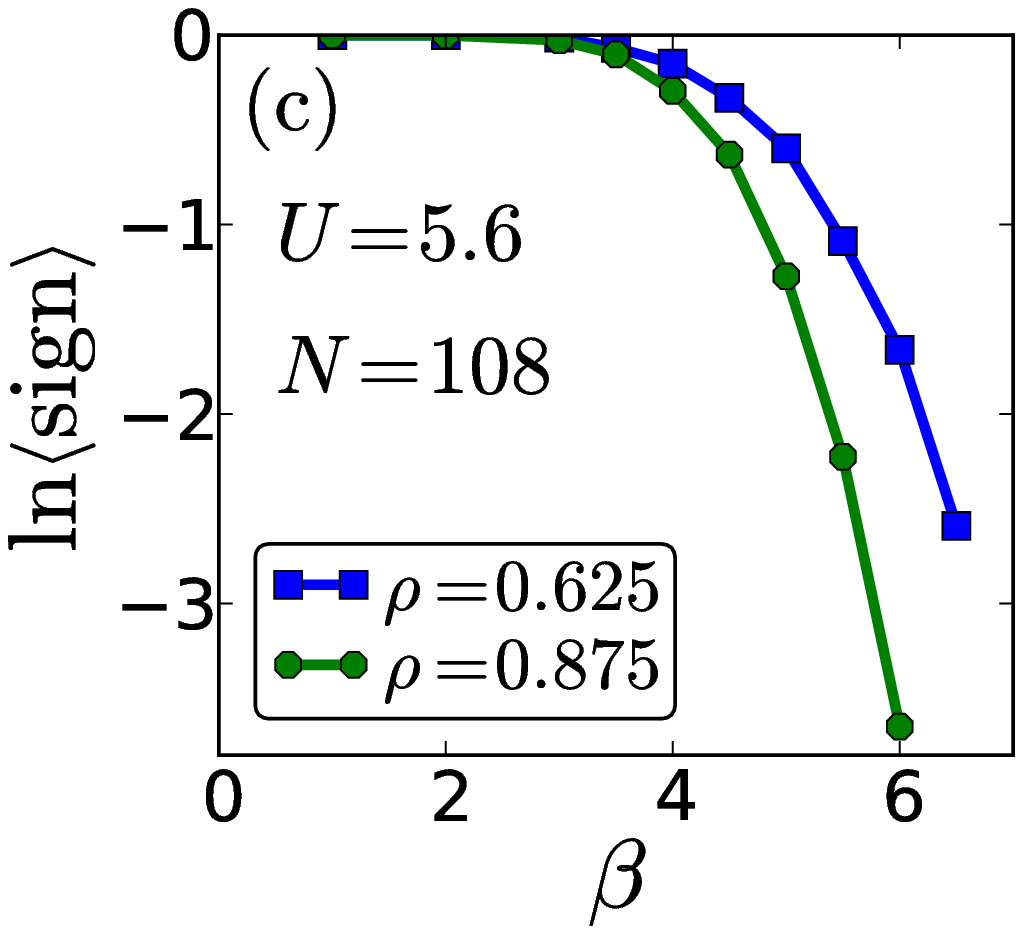, width=1.6in,angle=-0,clip}
\caption{(Color online) Same as Fig.~\ref{fig:chainU4beta8} but for the Lieb lattice
and with different parameters, which are noted in the panels.}
\label{fig:LiebU5.6beta6}
\end{figure}

We consider first the ``Lieb lattice" which consists of an underlying square array of ${\cal A}$ sites with additional two-fold coordinated ${\cal B}$ sites on each bond (see the left panel of Fig.~\ref{fig:Liebandonefifth}). This structure is a more chemically realistic multiband representation of the CuO$_2$ planes of the cuprate superconductors, with the Cu atoms forming the square array and bridging O atoms. The relevant filling of such a three-band model for the cuprates consists of one hole per CuO$_2$ unit cell, that is, well away from half-filling. Indeed, a realistic model of the cuprates would incorporate a substantial energy $\epsilon_{\rm pd}$ which represents the additional cost of holes to occupy an O $p$-orbital compared to a Cu $d$-orbital.

Nevertheless, the half-filled case, with three fermions per unit cell and $\epsilon_{\rm pd}=0$, has considerable interest: Lieb has given\cite{lieb89} a rigorous demonstration of a ferrimagnetic ground state in this situation.  The crucial observation is that the numbers of sites on the ${\cal A}$ and ${\cal B}$ sublattices ($N_{\cal A}$ and $N_{\cal B}$, respectively) are unequal.  Lieb showed that for any bipartite lattice with $N_{\cal B} > N_{\cal A}$ there is a `flat band' with $N_{\cal B} - N_{\cal A}$ zero energy eigenstates\cite{footnote3}.  A recent DQMC study has explored the attractive Hubbard model in this geometry~\cite{iglovikov14}.  The presence of a flat band was found to have important effects on physical properties like the local moment and pairing correlations.  Figure~\ref{fig:LiebU5.6beta6}  examines the sign problem for the repulsive case.  No qualitative difference is discernible from the square lattice.  Indeed, the sign shows no signal whatsoever as it passes  through $\rho=2/3$, the filling which corresponds to entry into the flat band.

\begin{figure}[t]
\epsfig{figure=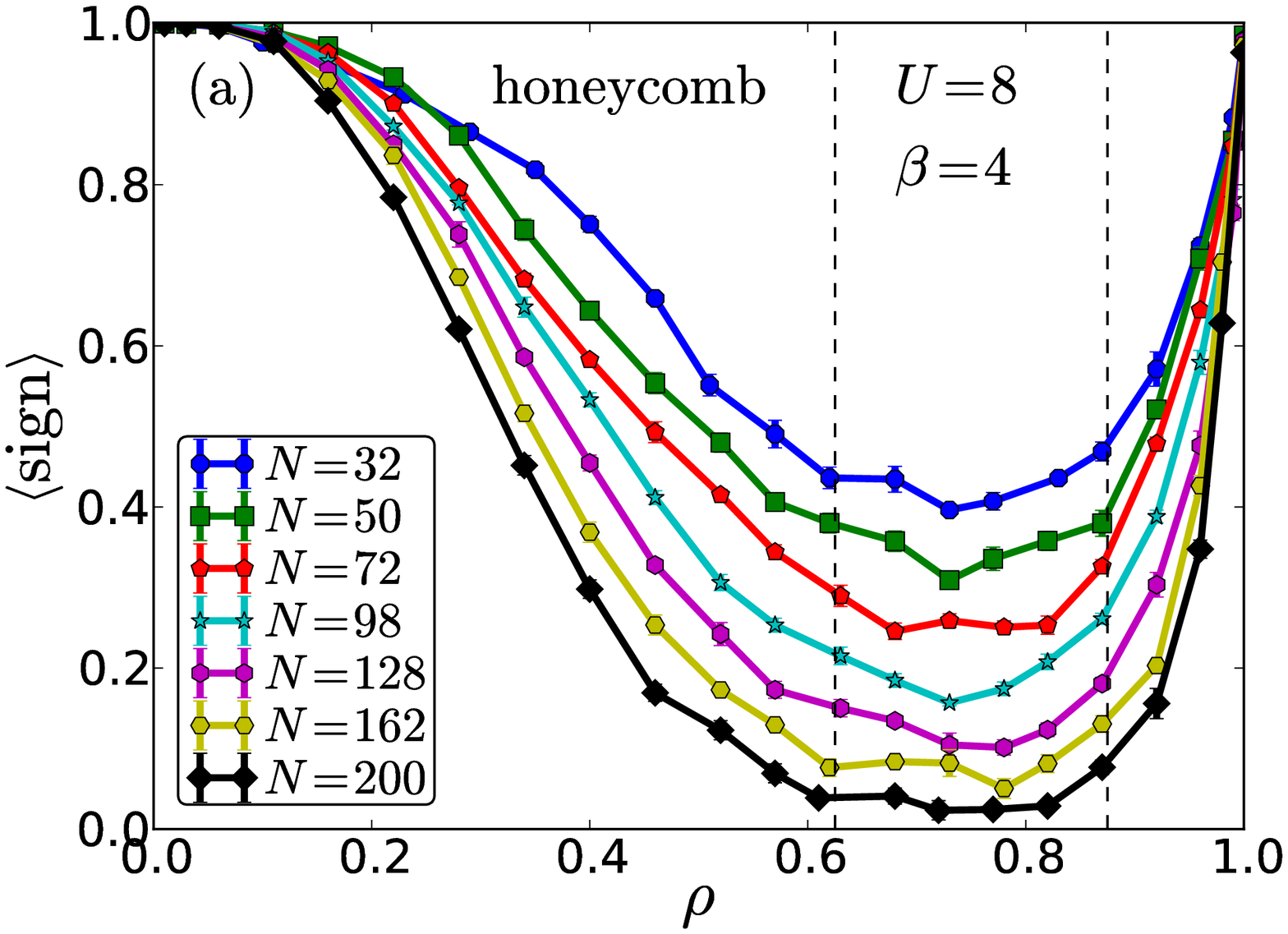, width=3.3in,angle=-0,clip}
\epsfig{figure=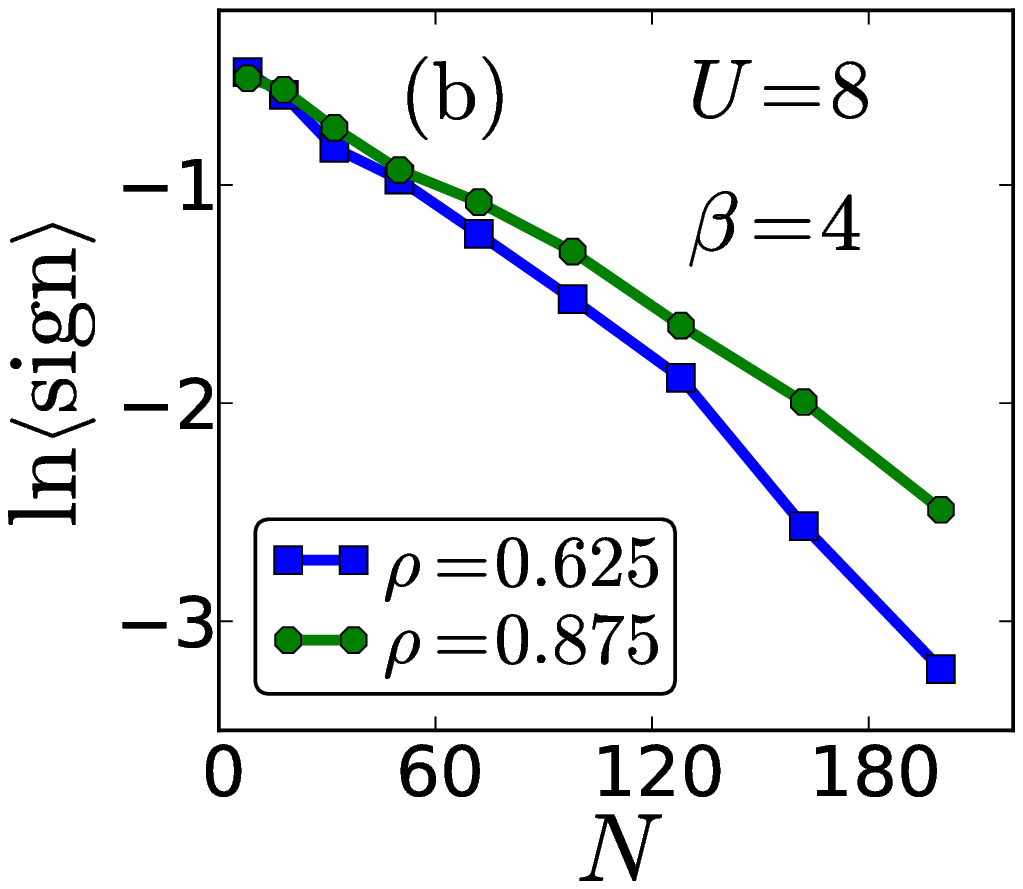, width=1.6in,angle=-0,clip}
\epsfig{figure=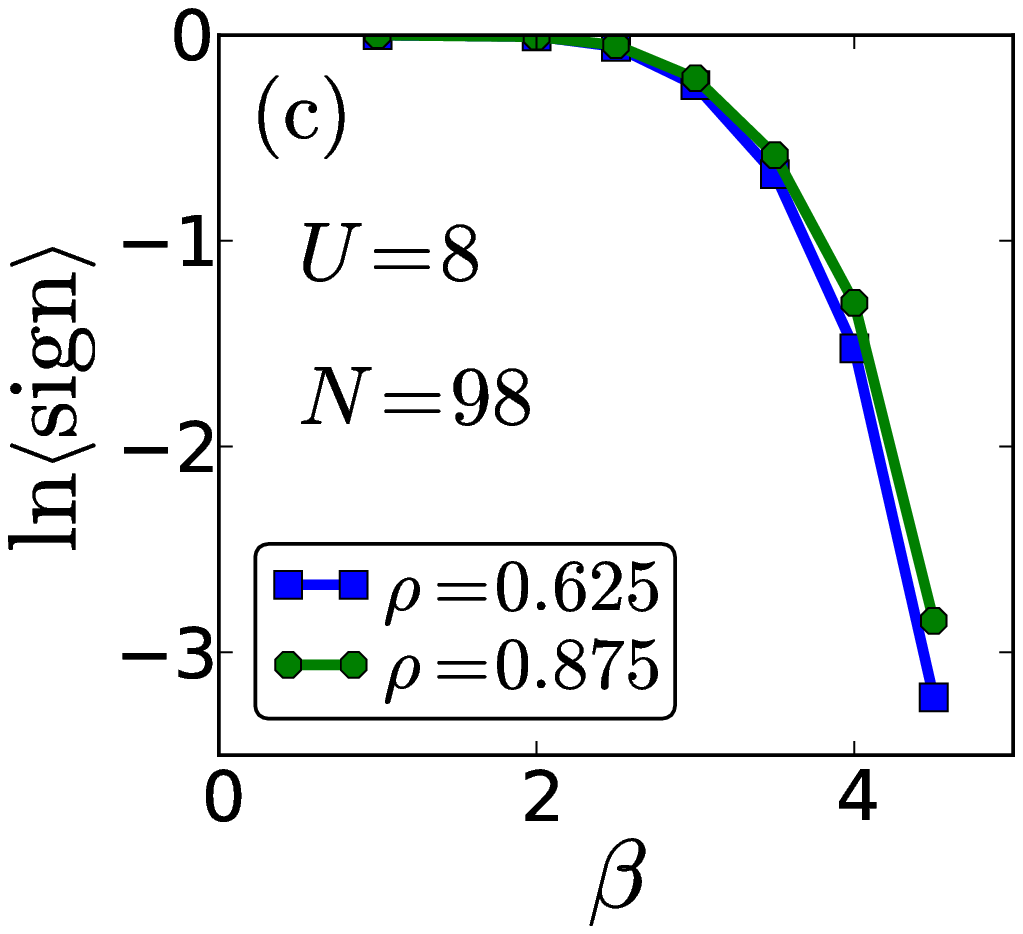, width=1.6in,angle=-0,clip}
\caption{(Color online) Same as Fig.~\ref{fig:chainU4beta8} but for the honeycomb geometry
and with different parameters, which are noted in the panels.}
\label{fig:honeycombU8beta4}
\end{figure}

The honeycomb lattice is another bipartite lattice we study here. It has an interesting semi-metallic density of states which vanishes linearly at $\omega \rightarrow 0$.  Like the hypercubic lattices, it has $N_{\cal A} = N_{\cal B}$. Figure \ref{fig:honeycombU8beta4} exhibits the average sign in the usual array of panels.  $\langle S \rangle$ is a bit reduced in densities $\rho \lesssim 0.6$ compared  to the other bipartite lattices, but otherwise behaves in a manner rather similar to them.

\begin{figure}[t]
\epsfig{figure=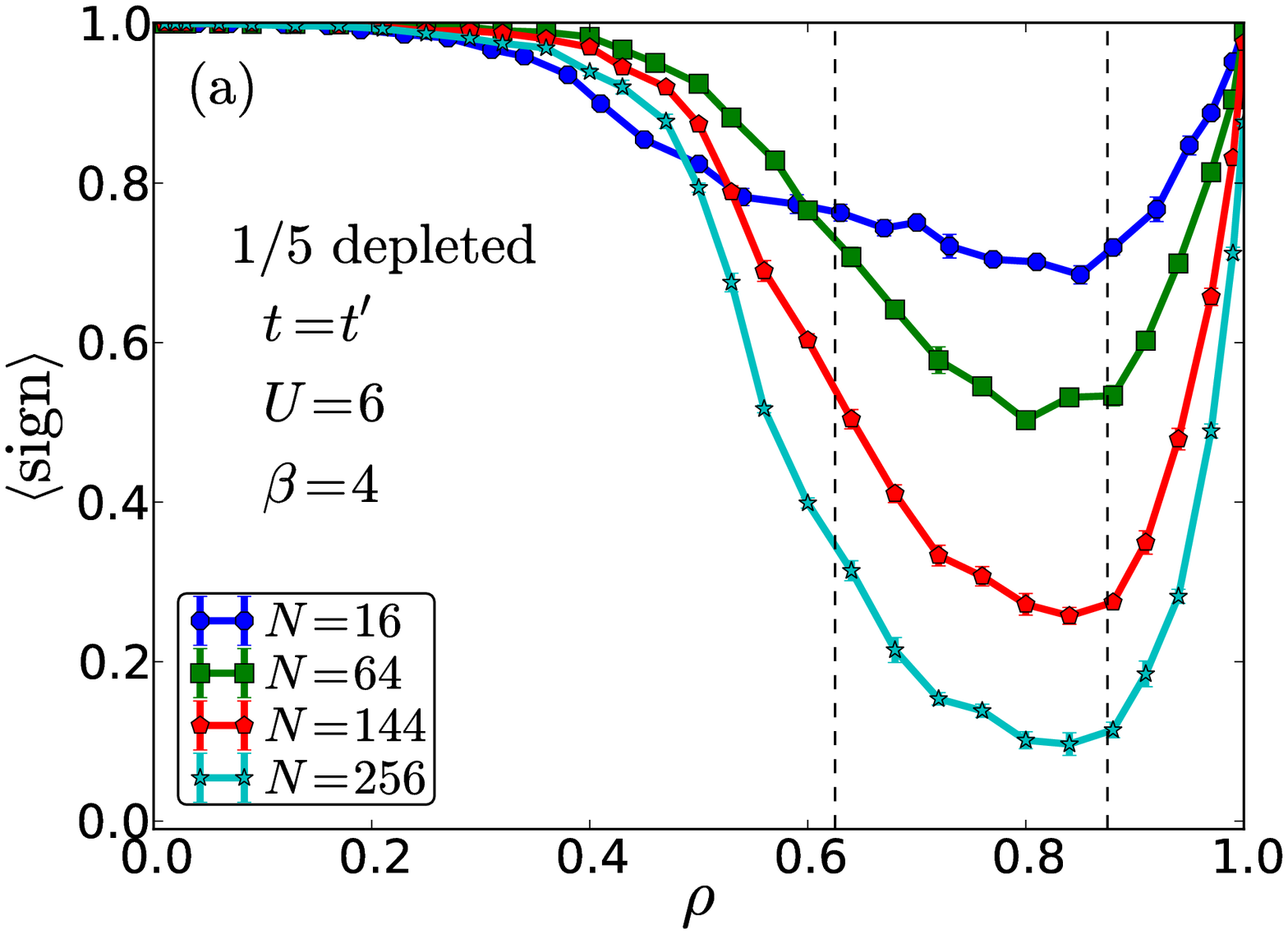, width=3.3in,angle=-0,clip}
\epsfig{figure=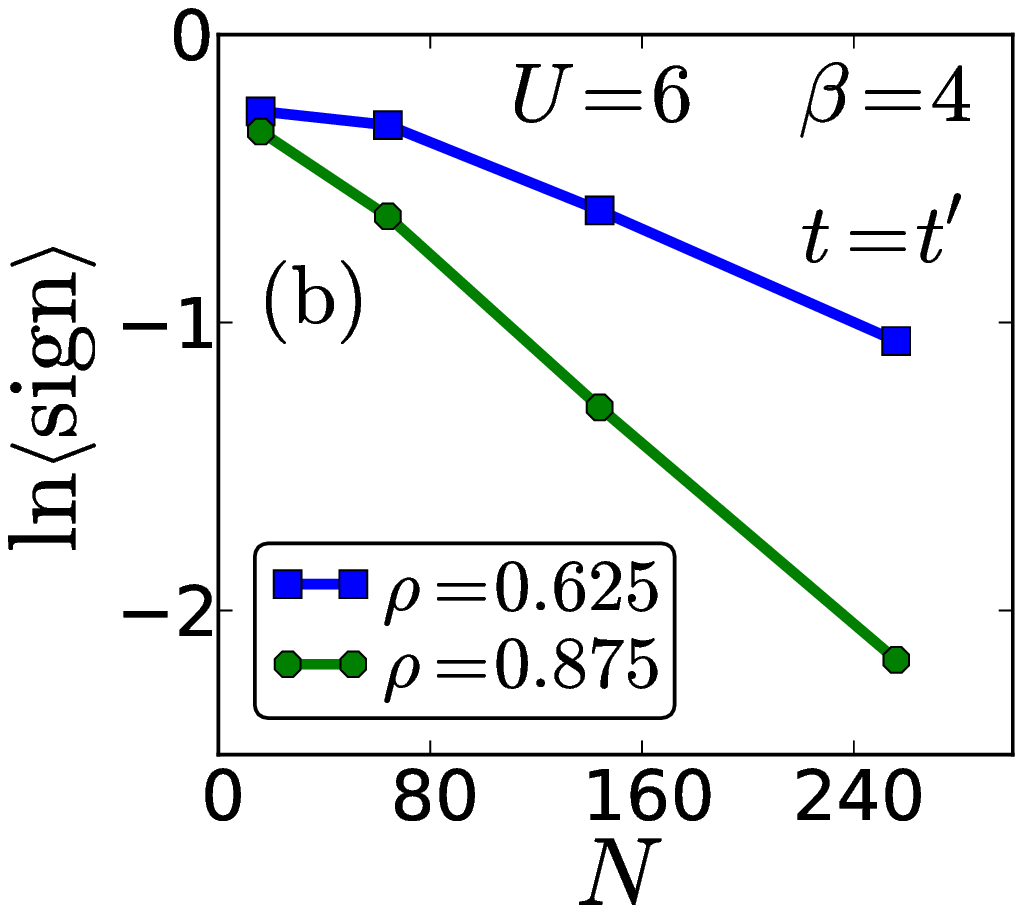, width=1.6in,angle=-0,clip}
\epsfig{figure=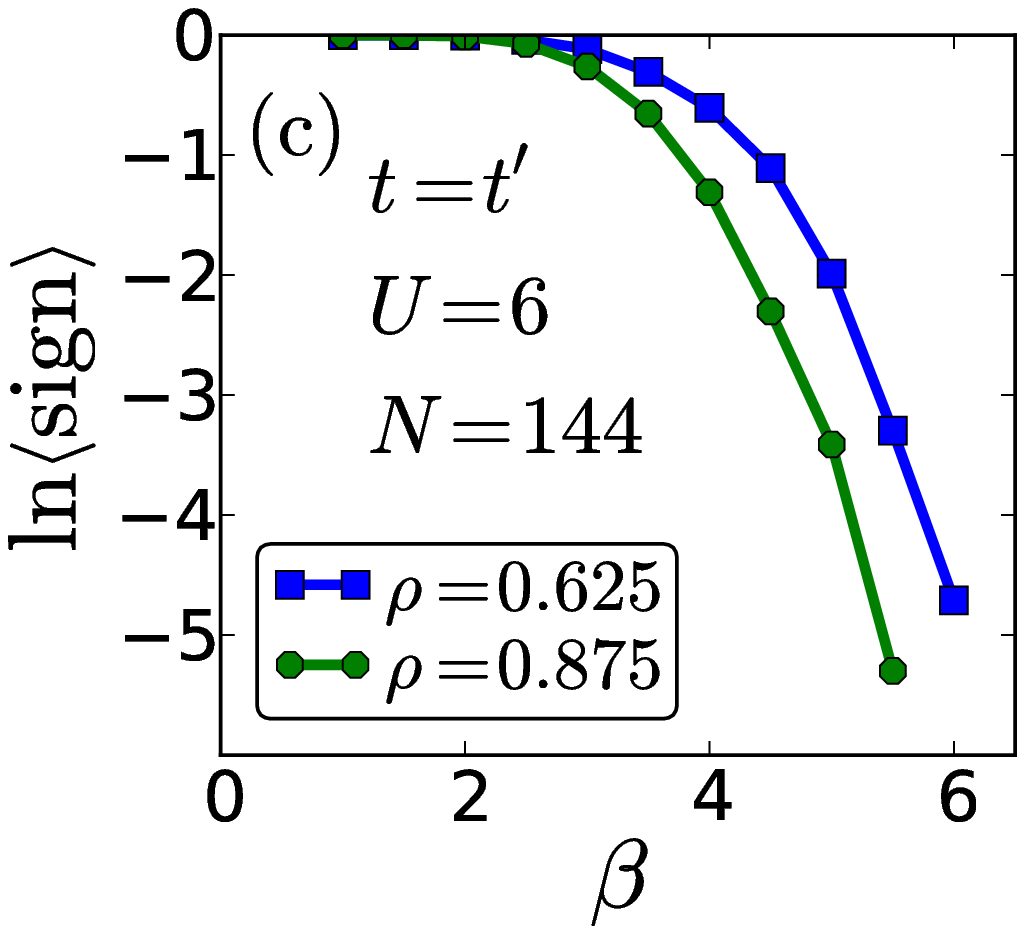, width=1.6in,angle=-0,clip}
\caption{(Color online) Same as Fig.~\ref{fig:chainU4beta8} but for the 1/5-depleted square lattice with
equal inter- and intra-plaquette hopping amplitudes and different parameters for $U$, $\beta$ and $N$,
which are noted in the panels.}
\label{fig:onefifth_1}
\end{figure}

\begin{figure}[b]
\epsfig{figure=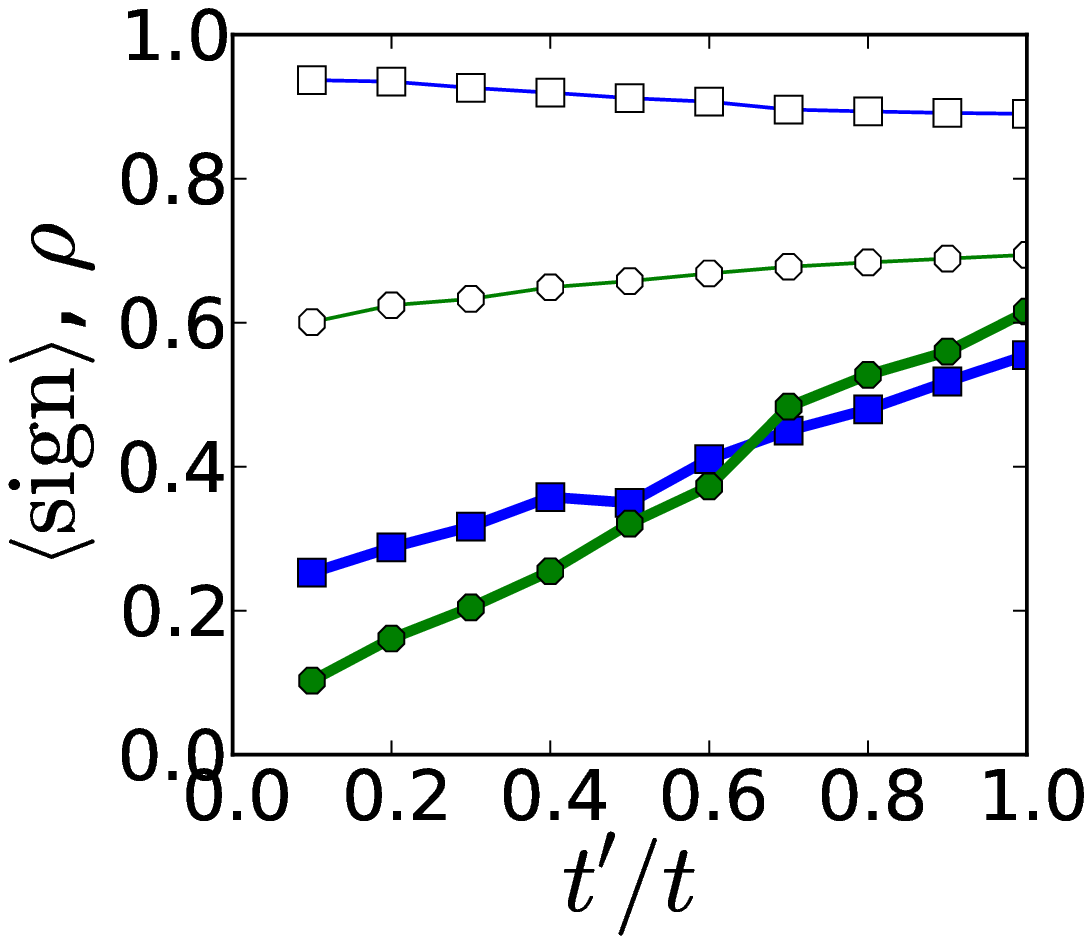, width=1.85in,angle=-0,clip}
\epsfig{figure=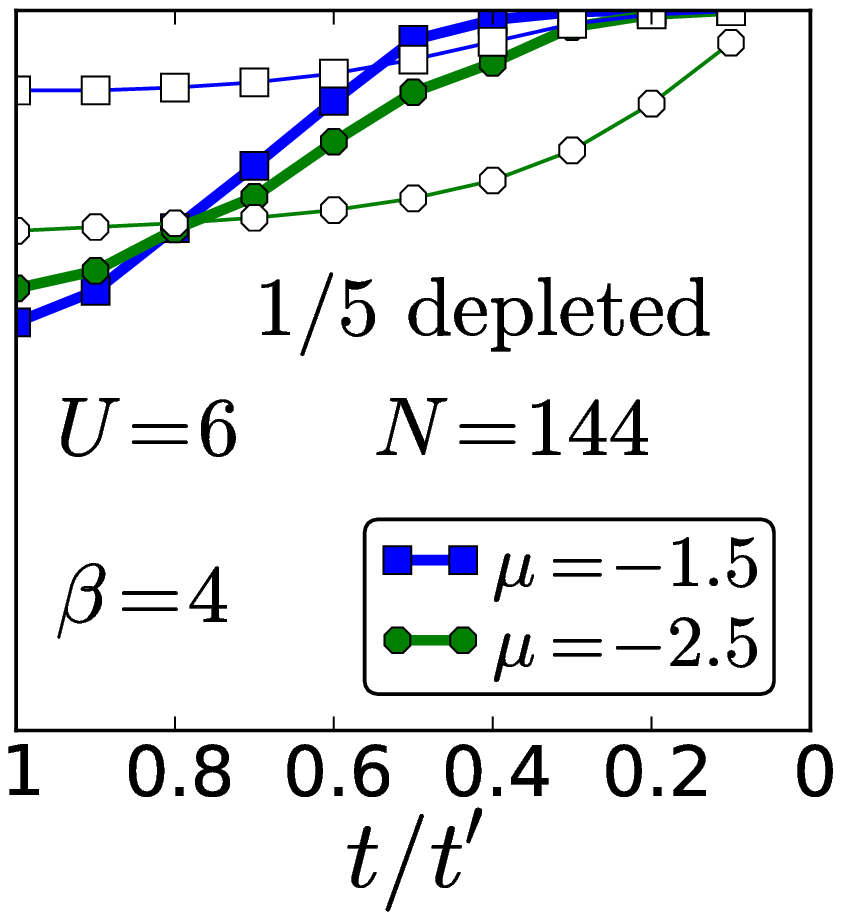, width=1.48in,angle=-0,clip}
\caption{(Color online) Evolution of the average sign on the 1/5-depleted square lattice as the ratio of the inter- and intra-plaquette hopping amplitudes ($t'/t$) increases. The results are obtained for two different fixed chemical potentials, $\mu=-1.5$ and -2.5. The filled symbols show the average sign and the empty symbols denote the evolution of the average density $\rho$. Note that $\rho$ varies slowly in most of the range of $t'/t$ and only increases significantly towards one for $t/t'\lesssim 0.5$. The bandwidth is kept fixed at 6 for all values of $t'/t$.}
\label{fig:onefifth_2}
\end{figure}

Our final bipartite geometry is a 1/5-depleted square lattice.  This is a cousin of the Lieb lattice, in that it can be regarded as a square lattice with 1/4 (rather than 1/3) of the sites removed, and is the geometry appropriate to the magnetic V atoms in CaV$_4$0$_9$ (see the right panel of Fig.~\ref{fig:Liebandonefifth}).  As with the Hubbard model on a Lieb lattice, this model exhibits interesting magnetic orderings.  In particular, at half-filling and $U=6$, as the ratio $t'/t$ of the inter- to intra-plaquette hopping is increased, one goes from a plaquette singlet phase to a phase with antiferromagnetic long-range order at $(t'/t)_{c_1} \approx 0.7$ and then to a dimer singlet phase where long-range order is again absent at $(t'/t)_{c_2} \approx 1.3$.  Figure~\ref{fig:onefifth_1} shows the doping dependence of the average sign for the 1/5-depleted square lattice for $t'=t$, which corresponds to the ordered phase at $\rho=1$.

Figure~\ref{fig:onefifth_2} gives the dependence on $t'/t$ for two fixed chemical potentials which correspond to $\rho \approx 0.9$ and $0.6 \lesssim \rho \lesssim 0.7$.  Although the density is varying a bit with $t'/t$, there is a steady decrease in $\langle S \rangle$ as $t'/t$ decreases. The dimer singlet phase at large $t'/t$ has a well-behaved sign, while the plaquette singlet phase has a much smaller average sign, presumably as a consequence of the fact that the sign problem of an isolated $2\times2$ plaquette is much worse than a dimer.

\subsection{Triangular and Kagome Lattices}


\begin{figure}[t]
\epsfig{figure=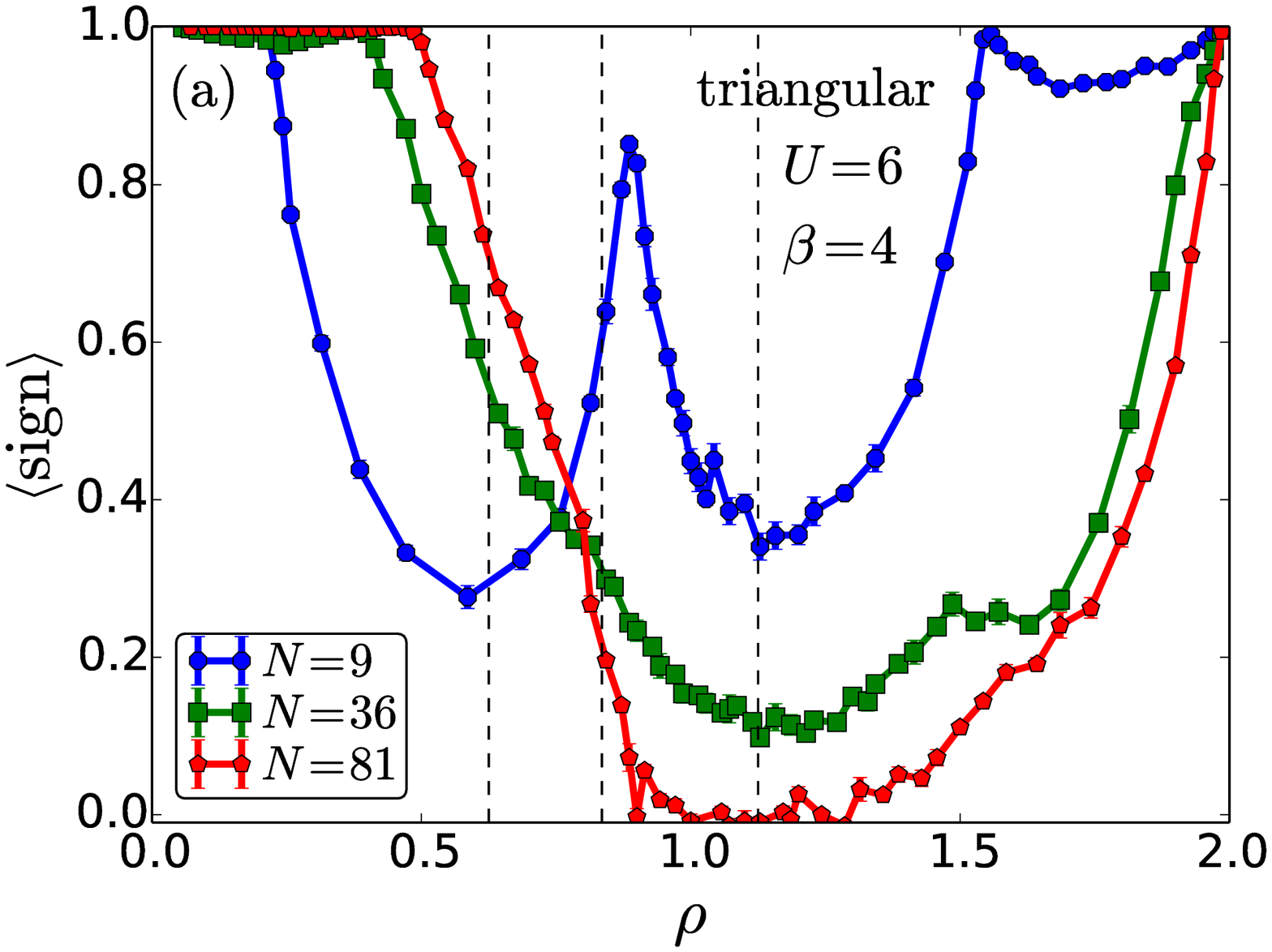, width=3.3in,angle=-0,clip}
\epsfig{figure=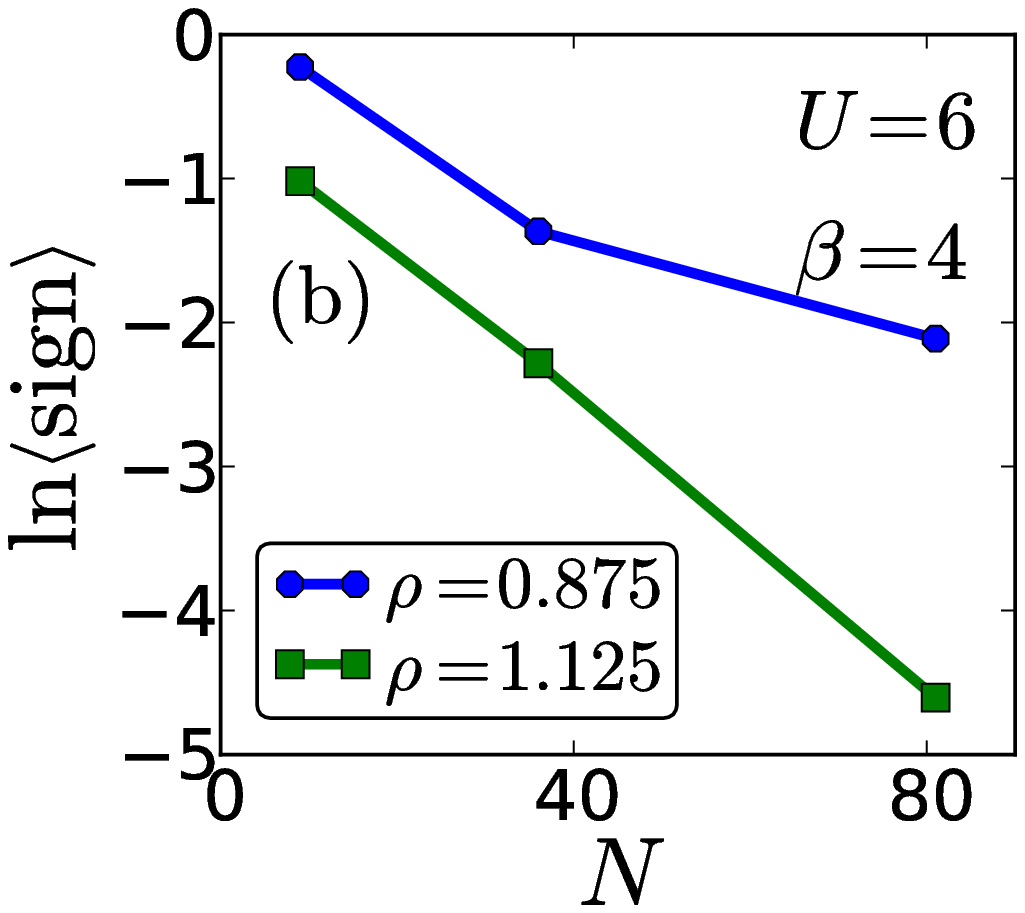, width=1.6in,angle=-0,clip}
\epsfig{figure=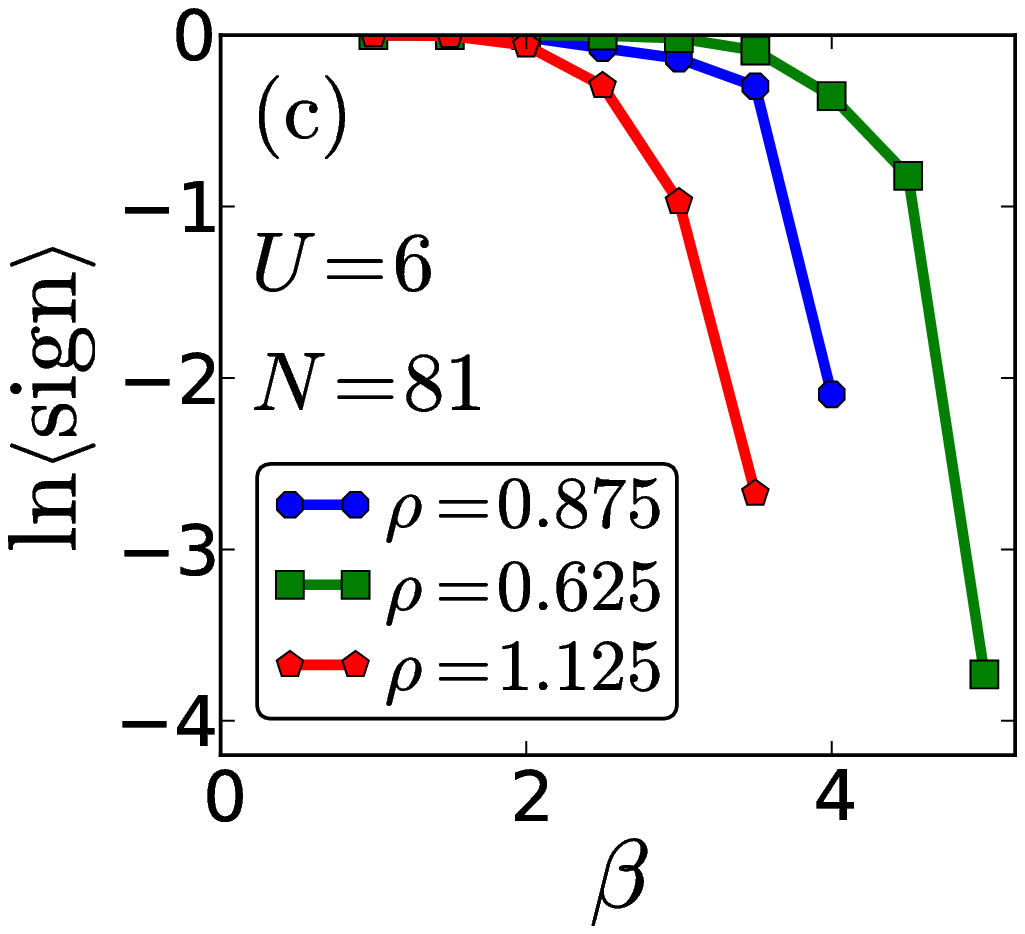, width=1.6in,angle=-0,clip}
\caption{(Color online) Same as Fig.~\ref{fig:chainU4beta8} but for the triangular geometry
and with different parameters, which are noted in the panels.}
\label{fig:triangleU6beta4}
\end{figure}


\begin{figure}[t]
\epsfig{figure=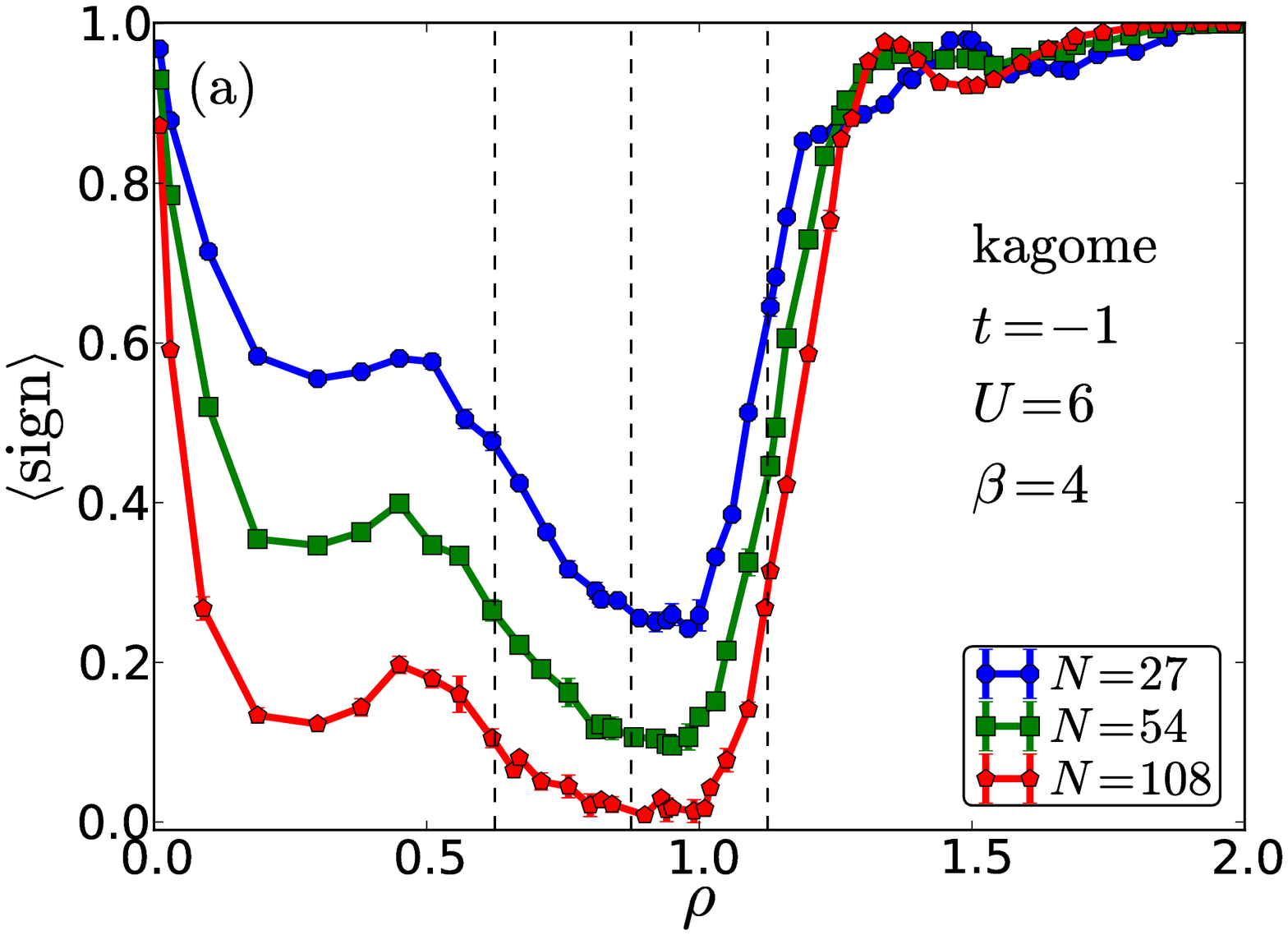, width=3.3in,angle=-0,clip}
\epsfig{figure=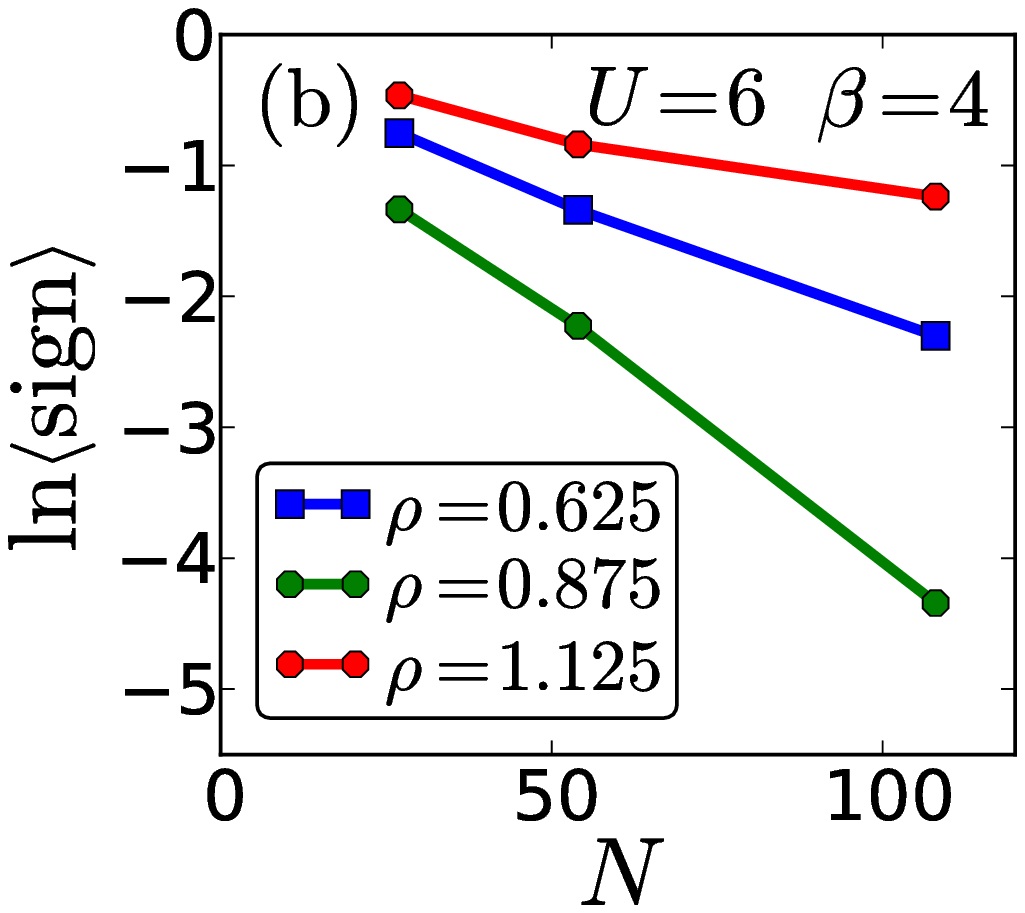, width=1.6in,angle=-0,clip}
\epsfig{figure=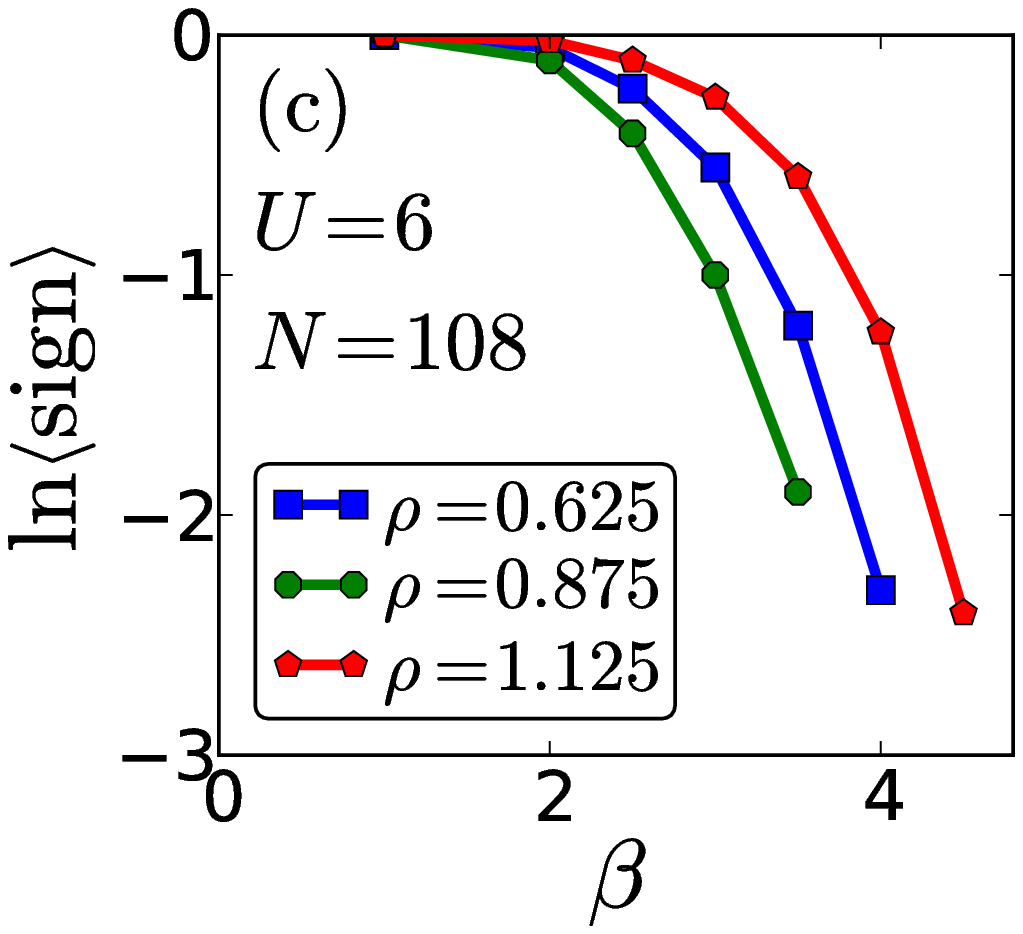, width=1.6in,angle=-0,clip}
\caption{(Color online) Same as Fig.~\ref{fig:chainU4beta8} but for the Kagome geometry
and with different parameters, which are noted in the panels.}
\label{fig:kagomeU6beta4}
\end{figure}

We conclude our survey of lattice geometries with two non-bipartite cases:  the triangular and Kagome lattices.  Since there is no PHS, we expect there will be a sign problem at all densities, including half filling. Moreover, we must present data over a full range of fillings, $0 \leq \rho \leq 2$. The sign of the hopping is also relevant for these structures. Our choices are $t=1$ for the triangular lattice and $t=-1$ for the Kagome lattice.  In terms of the density of states, these choices mean that the DOS is nonzero in the ranges [-6, 3] and [-2, 4] for the triangular and Kagome geometries, respectively.

Figure \ref{fig:triangleU6beta4} shows results for the triangular lattice.  Shell structure is evident for the smallest cluster ($N=9$), however, as with the square lattice, is absent for larger lattices ($N=36,81$).  The most marked difference from the bipartite cases is that $\langle S \rangle$ is not pinned at one for $\rho=1$, but otherwise the behavior of $\langle S \rangle$ is quite similar to the previous cases. The same is true of the Kagome lattice in Fig.~\ref{fig:kagomeU6beta4}, except that there is a persistent bump in $\langle S \rangle$ for $\rho$ slightly less than quarter filling. (A similar feature was noted for chains.) The structure in the average sign is somewhat more asymmetric about half filling than the triangular case.  One feature which does not seem to be shared with other geometries is the existence of an abrupt change in $\langle S \rangle$ appearing here at $\rho \approx 1.1$, so that it is well behaved for most densities $\rho \gtrsim 1.1$.  The scalings with the cluster size and $\beta$ of the average sign for these non-bipartite geometries, shown in the bottom panels of Figs.~\ref{fig:triangleU6beta4} and \ref{fig:kagomeU6beta4}, follow a similar pattern as for the other bipartite geometries.

\section{Further Analysis}

We now discuss possible patterns which emerge from these datasets. We focus on three areas: the role of the density of states, the contribution of individual spin components to the sign problem, and spatial entanglement. A fourth feature of $\langle S \rangle$, scaling in the vicinity of the PHS point, was discussed previously.

\subsection{Role of the $U = 0$ Density of States}

It seems plausible that the noninteracting density of states could play an important role in the sign problem.  In this subsection we make a few observations on that possibility.

The square and honeycomb lattices have quite dramatically different $U=0$ densities of states, especially near $\omega=0$ where the DOS diverges logarithmically for the square lattices, and vanishes linearly for the honeycomb lattice.  Yet, if we compare the behaviors of $\langle S \rangle$ as a function of filling in Figs.~\ref{fig:squareU6beta4} and \ref{fig:honeycombU8beta4}, we see little qualitative difference.  Both evolutions exhibit a rapid fall-off from $\langle S \rangle=1$ at the PHS point $\rho=1$, a broad minimum centered at $\rho \sim 0.8$, followed by a recovery to $\langle S \rangle=1$ in the dilute limit. The  differences in DOS are even more diverse among the other geometries studied here. However, the special features of the DOS, which could trend in completely opposite directions, appear to have little to no effect on the behavior of $\langle S \rangle$ here.

Indeed, we have noted already that, more generally, the different geometries and their associated distinct densities of states all share a qualitatively similar behavior of $\langle S \rangle$ with doping.  The only `unique' geometry was the one-dimensional case where, for example $\langle S \rangle$ did not recover to one at low densities.

\begin{figure}[t]
\epsfig{figure=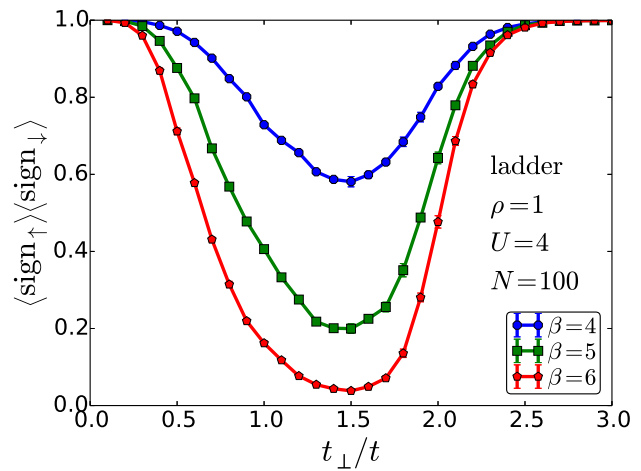, width=3.3in,angle=-0,clip}
\caption{The product of the expectation values of the signs of the individual spin up and spin down matrices is shown at half-filling as a function of the ratio of rung to on-chain hopping for the ladder geometry. ($\,\,\langle S_\uparrow \rangle = \langle S_\downarrow \rangle $ by symmetry.) The signs are well-behaved in the band insulating phase $t_\perp/t > 2$.}
\label{fig:spin_components_ladder}
\end{figure}

Most of the data we present are for interaction strengths $U$ at least four times the fermion hopping $t$, and hence, roughly speaking, at least half the bandwidth $W$.   These sorts of interaction strengths are the ones typically studied in examining magnetic, pair, and charge correlations in Hubbard models.  The conclusion of the observations above seems to be that $U \gtrsim W/2$ brings the system far enough from the $U=0$ limit that most features of the noninteracting DOS are no longer a controlling factors in the sign problem.  Apparently, the space and imaginary time fluctuations of the Hubbard-Stratonovich field ${\cal X}(i,t)$, whose effect on the fermions increases with $U$ through the parameter $\lambda$ of Eq.~\ref{eq:dhs}, smear out effects of the $U=0$ energy levels on $\langle S \rangle$. This appears consistent with a comparison of Fig.~\ref{fig:squareU6beta4}(a) of this paper with Fig.~10(a) in Ref.~\onlinecite{white89} for $4\times 4$ Hubbard lattices. With $\beta U$ constant, the sharpness of the feature near $\rho=0.6$ is significantly reduced when $U$ is increased from 4 to 6.

This lack of dependence on the DOS is the case even for a flat band, where the very large delta function in the DOS might have been expected to have an especially discernible impact on $\langle S \rangle$.  However, in the case of the Lieb lattice (Fig.~\ref{fig:LiebU5.6beta6}), $\langle S \rangle$ behaves completely smoothly through the edge of the flat band at $\rho=2/3$. We do note that broad features in the $U=0$ DOS do appear to have some correlation with the behavior of $\langle S_{\uparrow} \rangle = \langle S_{\downarrow} \rangle$. For example, for the chain, square, and cubic lattices, we observe that $\langle S_{\sigma} \rangle$ tends to be smaller when a ``smoothed" DOS is larger (not shown). Behavior near half-filling for bipartite lattices is additionally mediated, as discussed previously, by the fact that at half-filling $\langle S \rangle = 1$ by symmetry.

\subsection{Spin Components}

\begin{figure}[t]
\epsfig{figure=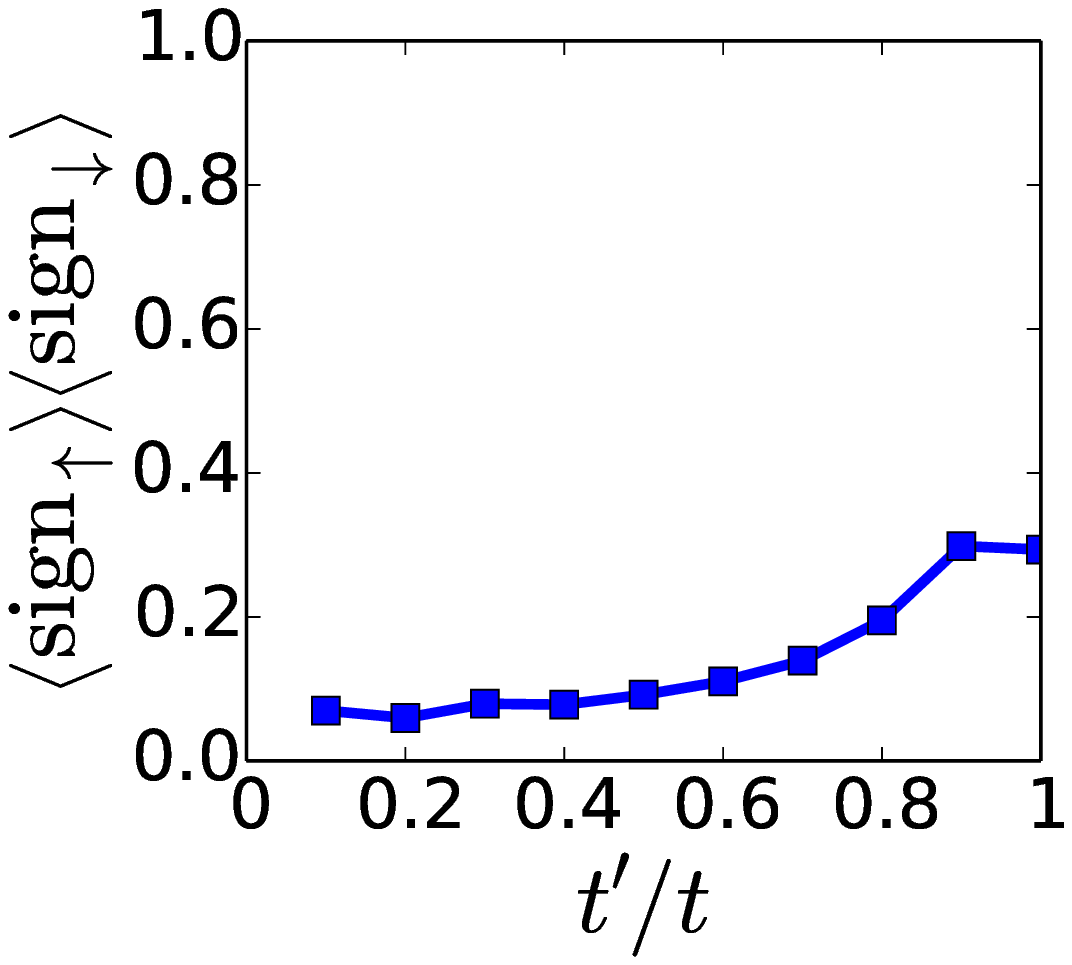,width=1.85in,angle=-0,clip}
\epsfig{figure=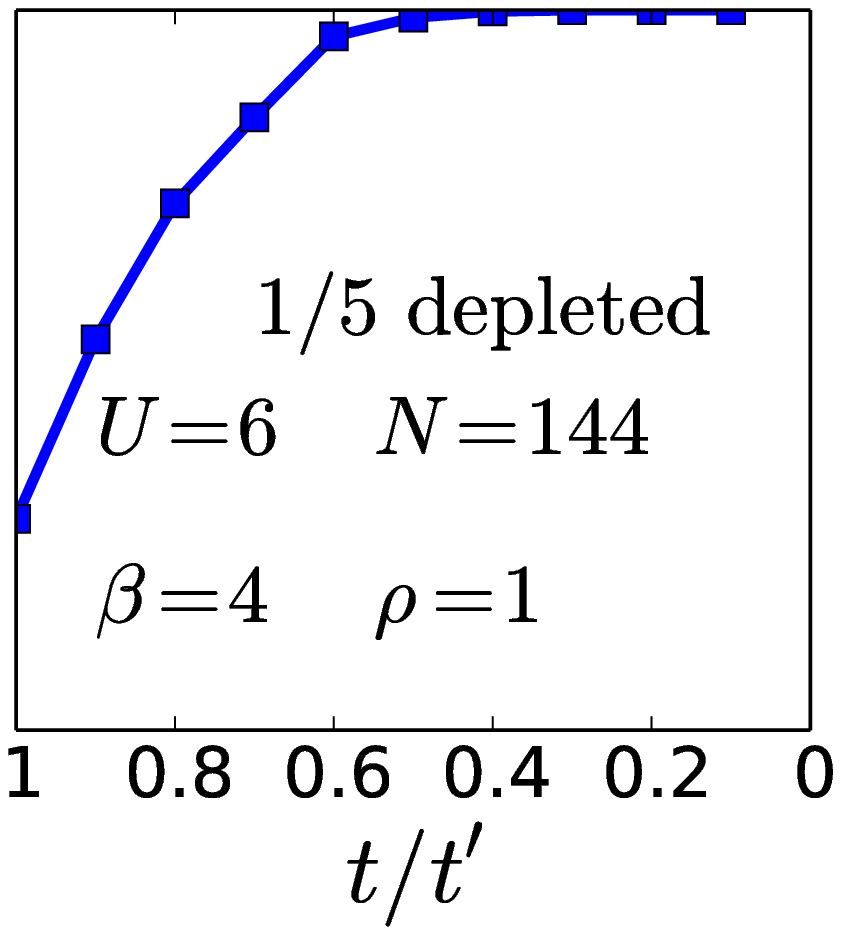, width=1.48in,angle=-0,clip}
\caption{(Color online) The product of the average signs for each spin specie on the 1/5-depleted square lattice at half filling as a function of $t'/t$.}
\label{fig:onefifthhalffilling}
\end{figure}

Thus far, we have focused almost exclusively on the total sign $\langle S \rangle$. As commented in Sec.~IIB, assuming there is no `off-diagonal' term in the Hamiltonian, allowing the mixing of different spin species, the fermionic trace results in separate determinants for each $\sigma$. Here, we do a further analysis of the signs of the individual spin components. We have two goals:  first, we would like to examine the correlations between the up and down determinants for the same HS configurations. Second, as in the case of the ladder or the 1/5-depleted square geometry, the spin-resolved sign may offer some insight into the potential connection of the average sign to other observables and phase transitions that occur in the PHS regimes (half filling).

We start with the second objective. We consider a ladder geometry with intra-chain hopping $t$ and inter-chain (rung) hopping $t_\perp$.  The $U=0$ band structure is $\epsilon_\pm(k_x) = \pm \, t_\perp - 2t \, {\rm cos} \, k_x$, so that the ladder is a BI for $t_\perp/t >2$ at half-filling, and a metal for $t_\perp/t<2$.  Figure \ref{fig:spin_components_ladder} shows the product $\langle S_\uparrow \rangle \langle S_\downarrow \rangle$ as a function of $t_\perp/t$ for $\rho =1$ (where $\langle S_\uparrow S_\downarrow \rangle=1$) and $U=4$ for $\beta=4, 5$ and $6$.  There seems to be some evidence that entering into the metallic phase at $t_\perp/t = 2$ coincides with an increase in the number of negative determinants.  The $t_\perp/t = 0$ limit is also interesting.  It corresponds to two decoupled 1D chains.  Evidently, even the individual determinants are free of negative signs at half-filling.

\begin{figure}[t]
\epsfig{figure=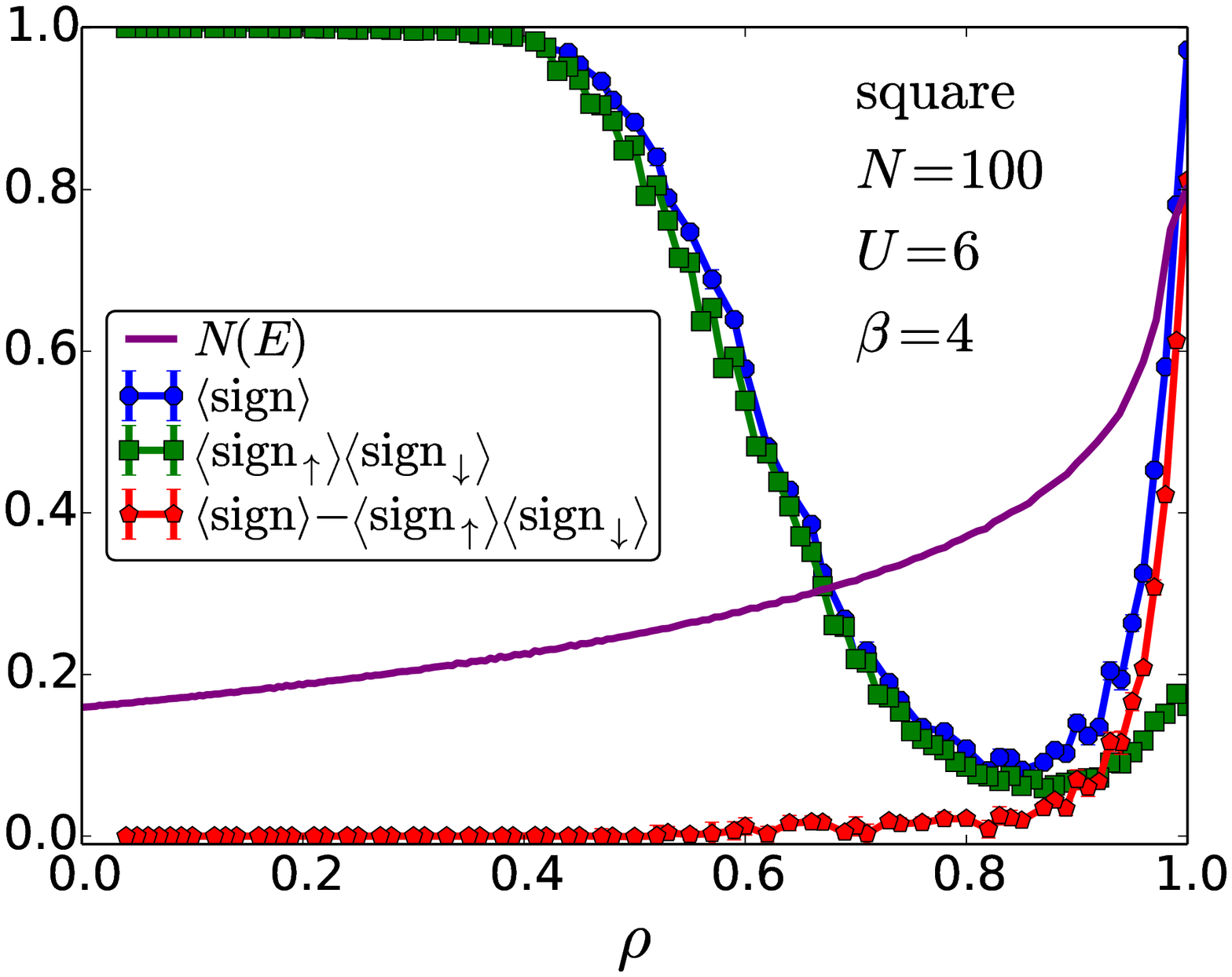, width=3.3in,angle=-0,clip}
\epsfig{figure=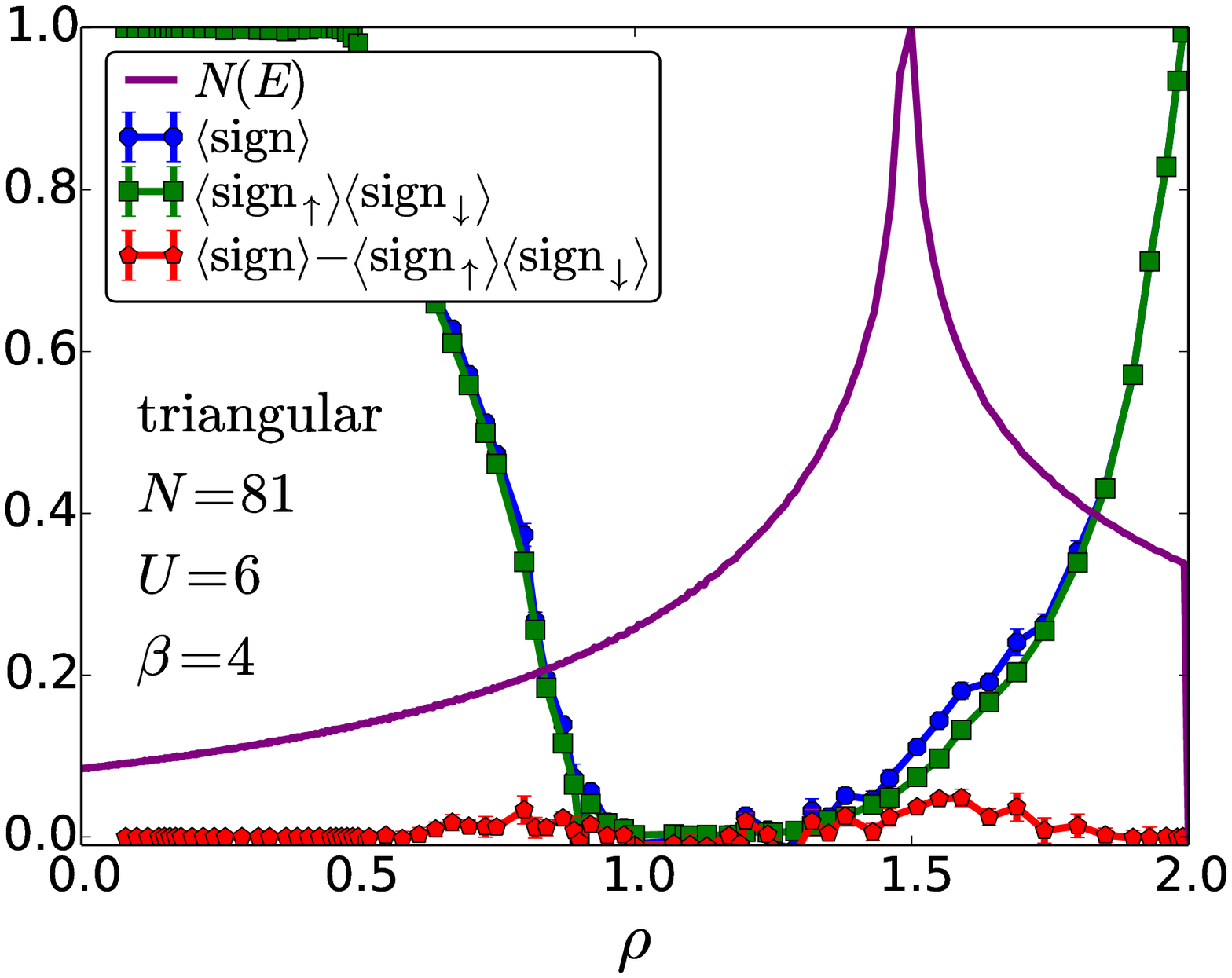, width=3.3in,angle=-0,clip}
\caption{Correlations between the average of the product of the up and down determinant signs and the product of the averages. Top panel: square lattice (PHS case). Bottom panel: triangular lattice (non-PHS case). The data indicate that, for the most part, the signs of the up and down determinants are independent, except at the PHS density $\rho=1$.}
\label{fig:spin_components_sq_and_triangle}
\end{figure}

Unlike for the ladder geometry, the product of the average signs for the two spin species does not display any signature at the (magentic) phase transitions for the 1/5-depleted square lattice. This can be inferred from Fig.~\ref{fig:onefifthhalffilling} where $\langle S_\uparrow \rangle \langle S_\downarrow \rangle$ at half filling is plotted vs $t'/t$ for $U=6$ and $\beta=4$. Similar to the total sign away from half filling (Fig.~\ref{fig:onefifth_2}), this quantity also decreases from one in the large $t'/t$ decoupled dimers region to near zero in the small $t'/t$ plaquette region. There are two quantum phase transitions\cite{khatami14} for this value of the interaction at $t'/t\sim 0.65$ and $t/t'\sim 0.77$. However, the average signs vary smoothly around these values with no special feature (at this temperature) that can be attributed to the phase transitions in the ground state.

With this in mind, we analyze the correlations between the up and down average signs for the square and triangular lattices in Fig.~\ref{fig:spin_components_sq_and_triangle}. In the former geometry, at half-filling, because of PHS there is a perfect correlation of the up and down signs, i.e., $\langle S \rangle=1$, which differs substantially from $\langle S_\uparrow \rangle \langle S_\downarrow \rangle$. Correlations remain in the neighborhood of the PHS point.  However, for densities $\rho \lesssim 0.8$ (and, by symmetry, $\rho \gtrsim 1.2$), $\langle S \rangle = \langle S_\uparrow \rangle \langle S_\downarrow \rangle$ to a high degree of accuracy, indicating that the signs of the spin up and spin down matrices for given HS configurations are essentially independent. It is interesting that the density at which correlations disappear roughly coincides to where the sign problem is the worst.

We expect the absence of the PHS to alter the relationship between the total sign and its individual components at half filling.  This issue is examined in the triangular lattice results in the bottom panel of Fig.~\ref{fig:spin_components_sq_and_triangle}.  The determinant signs are generally independent over the entire density range, with $\langle S \rangle = \langle S_\uparrow \rangle \langle S_\downarrow \rangle$, except for a small region around three quarter filling, $\rho \approx 1.5$.  This is evidently related to the bumps in $\langle S \rangle$ already noted in Fig.~\ref{fig:triangleU6beta4}, and provides additional insight into that phenomenon: The increase in $\langle S \rangle $ is associated, at least in part, with a larger correlation of the two determinant signs. Interestingly, the latter seems to be taking place close to where the peak in the DOS is, suggesting a possible connection, however, a similar but less prominent bump also appears around $\rho\sim0.7$, where there is no feature in the DOS.

\subsection{Spatial Entanglement}

Our data have supported the exponential decay of the sign with $N$ (Fig.~\ref{fig:cubicU5beta4.5} for the cubic lattice is inconclusive, perhaps due to the shorter lengths of the sides). Here, we point out that there is a limit in which this phenomena can be rigorously established. Specifically,  we observe that the sign must decay exponentially with $N$ for lattices consisting of decoupled clusters (by, for example, setting the inter-cluster hopping $t_{ij}$ to zero). Clearly the average sign of the entire system will be the product of the signs of the subclusters.  If the subclusters are identical, an exponential decrease of the average sign with system size (number of subclusters) trivially follows. One can then ask what is the effect of linking the clusters.  If the connection is through non-zero inter-cluster hopping, a `world-line' picture would generally suggest the sign should get even smaller, as the fermion paths are allowed additional opportunity to exchange.  In fact, precisely this would happen as a lattice of decoupled chains in the $x$ direction with zero hopping in the $y$ direction, $t_y=0$, is converted into a 2D lattice by turning on $t_y$. The $t_y=0$ limit has no significant sign problem in world-line approaches, whereas large $t_y$ does.

In DQMC, the opposite can occur: the linking (`entanglement') of spatial clusters mitigates the sign problem.  This can be seen in several Hubbard Hamiltonian geometries which contain decoupled clusters as a limiting case.  The simplest is the ladder geometry where $t_\perp=0$ corresponds to two independent chains.  Other examples are the `plaquette' model studied by Scalapino\cite{scalapino96}, Kivelson,~\cite{tsai08} and others,~\cite{doluweera08,baruch10,chakraborty11,karakonstantakis11, ying14} as a description of superconductivity arising from pair binding on $2\times 2$ plaquettes, and the 1/5-depleted square lattice~\cite{taniguchi95,troyer96,khatami14} for which the results in the left panel of Fig.~\ref{fig:onefifth_2} show the improvement of the sign by increasing the inter-plaquette hopping $t'$ from zero.

\section{Conclusions}

The sign problem remains one of the fundamental challenges in computational physics.  This paper has focused on the sign problem in determinant quantum Monte Carlo.  Our goal has been to bring together data for a collection of geometries (hypercubic, ladder, depleted square, Lieb, honeycomb, Kagome, and triangular lattices) and parameter (temperature, interaction strength, and density) ranges that are, at present either not available or, at best, scattered through the literature.  There are other cases one could study. However, this extensive set already enables us to make some general inferences about the sign problem, at least in the specific case of the DQMC, which may have other applications as well.

We first considered the general behavior of the sign $\left< S \right>$ as a function of inverse temperature $\beta = 1 / (k_B T)$ and of lattice size $N$.  Arguments for exponential behavior have been made in this regard involving the winding of world lines, in approaches where such paths are sampled in the simulation.  However, it has been noted that this reasoning does not necessarily transfer to the auxiliary field approach of DQMC, where it is difficult to see how to map the problem onto a sum over world line paths of an effective Hamiltonian with local interactions. Without such a theoretical basis, it is of interest to explore numerically the scaling of $\left< S \right>$ with $\beta$ and $N$ for a variety of cases, to determine the general behavior.

Previous work regarding $\beta$ used the Hubbard model on  $3\times 2$ lattices\cite{imada1} and  $4\times 4$ lattices\cite{loh90,dos_santos2003} finding an exponential decay of $\left< S \right>$ with $\beta$ for large $\beta$. Our data, on considerably larger spatial lattices and for a variety of different geometries, is also consistent with a large $\beta$ exponential decay, with a small $\beta$ regime throughout which $\left< S \right> \approx 1$, and eliminates `shell effects' seen in smaller lattices.

We next explored the scaling of $\left< S \right>$ with lattice size $N$. Previous work, comparing 2D $4\times 4$, $6\times 6$, and $8\times 8$ Hubbard lattices\cite{white89, dos_santos2003} and 3D $4\times 4\times 4$ and $6\times 6\times 6$ lattices\cite{dos_santos2003} was inconclusive.  In both the 2D and 3D cases, for example, for some parameter ranges $\left< S \right>$ was actually worst for the smallest systems studied. Using larger lattice sizes and differing geometries, our results are clearly consistent with an exponential decay of $\left< S \right>$ with $N$ for large $N$. In a few but not all cases, we also found a small $N$ regime in which $\left< S \right> \approx 1$, similar to what was seen for $\beta$. Data based on an examination of $\left< S \right>$ as a function of $U$ were consistent with a similar exponential decay of $\left< S \right>$ with $U$, for $U$ sufficiently large.

As is well known, the sign problem is `cured' in special particle-hole symmetric cases like the half-filled Hubbard model on a bipartite lattice, where the determinants always come in pairs which share the same sign. This paper presented a more general study of correlations between the signs. Near half-filling, for the square lattice, we found $\left< S \right> > \left< S_{\uparrow} \right> \left< S_{\downarrow} \right>$, where $\langle S_{\sigma} \rangle$ is the average sign of a single spin determinant, see Eq.~6. This indicates the development of strong correlations between the signs of the up and down determinants, 

A closely related issue is the precise form of the decay of $\left< S \right>$ close to the half-filled $\rho=1$ PHS point where $\left< S \right> = 1$.  We found a rapid decay, consistent with the form $\left< S \right> = e^{a |\rho - 1|}$ and that ``$a$"becomes increasingly large and  negative with $U$ and with $\beta$. This indicates a rapid loss of the positive correlations between up and down determinants as one moves away from half filling.

Lastly, we considered the effect of ``entanglement", where initially decoupled clusters were linked together with fermion hopping terms. This linking  increased $\left< S \right>$ improving the sign problem, in complete contrast to what one would expect in a ``world line" picture, where the additional windings available due to the extra hopping terms would typically be expected to worsen the sign problem.

Another significant feature of this work is that the geometries being studied have a wide range of noninteracting densities of states. A crucial conclusion of our work, evident in comparing the top panels of Figs.~2-7 is that, despite the wide variation in the DOS, on bipartite lattices, the behavior of the sign with density is almost ``universal" in the sense that it always falls rapidly away as the lattice is doped, attains a minimum in the vicinity of $\rho \approx 0.8$ and then recovers. This observation is of interest because of the rather different physics expected of strongly correlated electrons on these geometries. The only exception to this universality is seen in Fig.~\ref{fig:chainU4beta8}, for the one dimensional chain.

Overall, features in the $U = 0$ DOS associated with special behavior at an isolated energy seem to have little effect on $\left< S \right>$. This may be due to the ``smoothing out" of these features by the HS field fluctuations which increase with increasing $U$. There does, however, seem to be some indication that a larger value of a ``smoothed" DOS is associated with smaller average signs $\left< S_{\uparrow} \right> = \left< S_{\downarrow} \right>$ of the determinants of the matrices of individual spins.

Another interesting possibility for further exploration is that of a linkage between $\langle S \rangle$ and the spectral function at the Fermi level, $A(\omega=0)$. Here $A(\omega)$ is related to the time dependent Green's function via $G(\tau) = \int \, d\tau \, e^{-\omega \tau} A(\omega) \, / \, ( \, e^{\beta \omega} + 1 \,)$ and equals the density of states $N(\omega)$ in the non-interacting limit. The spectral function incorporates the effects of $U$ and hence, potentially, might correlate better with the sign. There are some hints that this is the case, for example, on a square lattice the $U=0$ density of states diverges at half-filling yet the low-temperature $A(\omega)$ vanishes for all $U$ and the sign behaves perfectly.  Indeed, the sign behaves well in a range of chemical potentials $\mu$ within the Mott-Slater gap.  One could imagine pursuing this possible connection more closely through computing the spectral functions for the various geometries and parameter values considered here.  However, this would be a major task beyond the scope of this paper, involving the measurement of the imaginary-time dependent Green's function and its analytic continuation to real space. In addition, because it is more  difficult to obtain $A(\omega)$, as compared  to the noninteracting density of states, it is unclear what the utility of the discovery of a connection between $A(\omega)$ and the sign would be.

Despite the data and interpretation presented here, the sign problem remains a big mystery. Of particular interest is the possible relation between $\langle S \rangle$ and the underlying physics of correlated electrons. For example, early in the development of DQMC, Hirsch pointed out\cite{hirsch85} a mapping between the spin-spin correlation function, a property solely of the Hubbard Hamiltonian itself, and a correlation function of the Hubbard-Stratonovich field ${\cal X}$. Since the fluctuations in ${\cal X}$ ultimately determine the sign problem, this suggests the possibility that the behavior of $\langle S \rangle$ might be related to some appropriate observable.

We end on a speculative note in this regard. The unfortunate coincidence (if it is a coincidence) that for the square lattice $\langle S \rangle$ is worst behaved very close to the most interesting `optimal' doping, where $T_c$ is largest in the cuprate superconductors, has often been opined. One thing we observe here is that this dip in $\langle S \rangle$ at $\rho \approx 0.8$ is absent in $d=1$ but appears already in the ladder geometry. Since Hubbard ladders appear to show signatures of $d$-wave pairing\cite{noack94} the possibility of a deep connection between the fermion sign and superconductivity in the Hubbard Hamiltonian remains a possibility.

\section*{ACKNOWLEDGMENTS}

We are very grateful to Richard Fye for useful discussions concerning this manuscript and the sign problem. This work was supported by the University of California Office of the President. This work used the Extreme Science and Engineering Discovery Environment under Project No. TG-DMR130143, which is supported by NSF Grant No. ACI-1053575



\begin{thebibliography}{100}
\bibitem{footnote3}

Since these eigenstates have degenerate eigenenergy, linear combinations can be taken so that they may either be viewed as plane wave states forming a flat band, or as spatially localized states.
\end{thebibliography}
\end{document}